\documentclass[a4paper,12pt]{article}

\usepackage{amsmath,amsfonts,amssymb,epsfig}

\numberwithin{equation}{section}

\usepackage{cancel}
\usepackage{cite}

\def\bra{\langle}
\def\ket{\rangle}

\def\tr{\mathrm{tr}}

\def\beq{\begin{equation}}
\def\eeq{\end{equation}}
\def\bal{\begin{align}}
\def\eal{\end{align}}

\def\Mm{\M_{\mathrm{mess}}}

\def\2b2[#1,#2][#3,#4]{\left( \begin{array}{cc} #1 & #2 \\ #3 & #4 \end{array} \right)}
\def\3b3[#1,#2,#3][#4,#5,#6][#7,#8,#9]{\left( \begin{array}{ccc} #1 & #2 #3 \\ #4 & #5 & #6\\#7&#8&#9\end{array} \right)}

\newcommand{\C}[1]{\mathcal{#1}}

\def\Mm{M_{\mathrm{mess}}}

\def\ov{\overline}

\author{K.~Benakli$^1$\footnote{kbenakli@lpthe.jussieu.fr} and M.~D.~Goodsell$^2$\footnote{mark.goodsell@desy.de}}
\date{}
\title{\vspace{-3cm}
\hfill{\small{DESY 10-044}}\\[2cm]
Dirac Gauginos, Gauge Mediation and Unification}
\begin{document}
\maketitle
\vspace{-1cm}
\begin{center}
\emph{$^1$Laboratoire de Physique Th\'eorique et Hautes Energies,  CNRS, UPMC Univ Paris 06
Boite 126, 4 Place Jussieu, 75252 Paris cedex 05, France \\
$^2$Deutsches  Elektronen-Synchrotron, DESY, Notkestra\ss e 85, 22607  Hamburg, Germany}
\end{center}
\abstract{We investigate the building of models with Dirac gauginos and perturbative gauge coupling unification. Here, in contrast to the MSSM, additional fields are required for unification, and these can naturally play the role of the messengers of supersymmetry breaking. We present a framework within which such models can be constructed, including the constraints that the messenger sector must satisfy; and the renormalisation group equations for the soft parameters, which differ from those of the MSSM. For illustration, we provide the spectrum at the electroweak scale for explicit models whose gauge couplings unify at the scale predicted by heterotic strings.}

\newpage

\section{Introduction}

The Large Hadron Collider (LHC) will soon investigate the terascale energy frontier and search for new interactions and particles. These are predicted by many extensions of the Standard Model, and in particular supersymmetric ones predict the existence of  many new particles.  The existence of supersymmetry can be related to the issue of unification of known interactions. From the top-down approach, supersymmetry seems to be an important ingredient in models of quantum gravity such as string theory. From the bottom-up approach, it allows to address the problem of the large hierarchy of between the electroweak  and the fundamental (quantum gravity) scale. The possibility of unifying  all interactions can then be considered as an important issue in model building of supersymmetric extensions of the Standard Model.

The prediction of unification of the gauge couplings at scale $M_U \sim 2 \cdot 10^{16}$ GeV, compatible with a GUT structure, can be considered as one of the main successes of the Minimal Supersymmetric Standard Model (MSSM)  \cite{Dimopoulos:1981yj}. It is based on the assumption that no new states carrying Standard Model gauge quantum numbers lies between the electroweak and unification scales. This desert scenario is abandoned when the origin of the MSSM soft masses is investigated. For instance, if they are explained as being due to gauge mediation \cite{Dine:1981za} (for a review, see for example \cite{Giudice:1998bp}),  messenger states carrying $SU(3)\times SU(2)\times U(1)_Y$ charges  are introduced at intermediate energies and modify the running of the gauge couplings. Unless an additional constraint is imposed, requiring that the messengers are chosen  in suitable representations (such as complete $SU(5)$ multiplets), unification is lost.

Upon the discovery of new fermions,  the question of their nature, Dirac or Majorana, can be raised. It is then legitimate to ask about the possible existence of models with Dirac gaugino masses, and if so, investigate their main features. To allow such  masses, the MSSM needs to be extended by extra adjoint representations, which we denote as DG-adjoints, that couple to the gauginos. The possibility of building such models has attracted some interest  in the past (see for example \cite{Fayet:1978qc, Polchinski:1982an, Hall:1990hq, Fox:2002bu, Nelson:2002ca,Antoniadis:2005em,Antoniadis:2006uj, Amigo:2008rc, Benakli:2008pg, Belanger:2009wf,Benakli:2009mk, Chun:2009zx}). A substantial motivation for such models is that, in contrast to Majorana gauginos, the supersymmetry breaking sector can preserve $R$-symmetry, allowing for a wider variety of breaking scenarios; we shall also assume this feature here. 

Here, we shall investigate the issue of their compatibility with  gauge coupling unification. In fact,  except for  the case of the $U(1)_Y$ bino, where the DG-adjoint is just a singlet superfield (that could  be identified with a generic modulus field \cite{Benakli:2009mk}), since the DG-adjoints are introduced at the electroweak scale, they will drastically modify the running of the gauge couplings.  Moreover, the running of the couplings is also modified by the messengers  introduced in order to induce the right soft masses in the framework \cite{Benakli:2008pg} which extends generalized gauge mediation \cite{Meade:2008wd}. In this work, we will show that the situation of models of Dirac gauginos can be improved as the messengers  can restore unification. In contrast to the MSSM, the messengers are not only allowed but needed for unification.

One of the main problems in models with  DG-adjoints is the tendency to create a Landau pole at an intermediate scale. In fact, the colour octet makes the $SU(3)$ coupling no longer asymptotically free, and any new coloured states will make it grow quickly in the UV. In order to avoid a Landau pole, we need to make the messenger masses as heavy as possible. On the other hand, it is this messenger scale  that appears to suppress the induced soft-masses, and to obtain sizable values for the latter, we need to keep the messengers light. We find that the best way to resolve this tension is to take the messenger masses to be intermediate between the unification and electroweak scales, and to use a combination of both $D$ and $F$ terms to generate the soft masses, the gaugino masses being dominated by the effect of the first, while the MSSM scalar masses are dominated by the effect of the second. 

In section 2, we will describe the content and the interactions in our minimal supersymmetric extension of the Standard Model with Dirac gauginos, which we denote DG-MSSM. We will exhibit the structure of the effective Lagrangian at the electroweak scale and point out the main differences with the MSSM, in particular the presence of the so called ``non-standard'' supersymmetry breaking terms. These   arise in combinations that  make them (super)-soft \cite{Fayet:1978qc,Polchinski:1982an,Nelson:2002ca}. An important issue is the fate of $R$-symmetry that is preserved by the soft terms explicitly computed. Since it is a global symmetry so must therefore be broken, and moreover a generalisation of chiral symmetry, we consider it to be natural that it is broken in the Higgs sector.  However, although we discuss some interesting possibilities in section \ref{Higgs}, we postpone to future work a comprehensive investigation of the  Higgs sector, and instead in common with many other gauge mediation models introduce explicit $\mu$ and $B_\mu$ terms. In order to obtain explicit examples for the spectrum at the electroweak scale, we assume that these are generated by some unspecified additional mechanism and keep their values small, such that if the same source generates $R$-symmetry breaking Majorana masses for the gauginos, these remain sub-leading.

The messenger sector is discussed In section 3. We introduce a superpotential describing the couplings of the messengers to the visible sector fields, as well as their couplings to the spurion fields that parametrise the breaking of supersymmetry by an unspecified secluded sector. We discuss the generic formulae for the soft masses induced either by  $D$-term, or by $R$-symmetric $F$-terms. If in both cases $R$-symmetry is supposed to be preserved at the leading order,  no Majorana masses are generated for the gauginos. The corresponding constraint on the superpotential parameters  is given in section 4. This section also summarizes many other constraints on the model's high energy parameters. Because  of the high values needed  for the messenger masses as well as for the $F$ and $D$ terms,   quite severe constraints on the model parameters arise in order to avoid tadpoles and generation of tachyonic masses for the adjoint scalars. We list these constraints in section 4. 

Because the messengers are quite heavy, the running of the induced soft-terms from the messenger scale down to the electroweak scale is very important and should be taken into account at leading order. The generic formulae for the evolution of the couplings, the standard and non-standard  soft terms are available in the literature \cite{Jack:1999ud,Jack:1999fa,Martin:1997ns}, and can be implemented for the specific case of the DG-MSSM. These are presented in section \ref{RENORMALISATION} and for more general models (allowing R-symmetry breaking terms) in appendix \ref{Appendix:RGEs}. The main observation is that the Dirac gaugino masses do not contribute to the one-loop running of the MSSM scalar masses other than the Higgs, in contrast to the Majorana ones which contribute to all.

The unification of Standard Model gauge couplings is illustrated in section \ref{UNIFICATION}. We deliberately choose the unification scale to match the heterotic string prediction \cite{Kaplunovsky:1987rp}, which allows unification of all known interactions. With our choice of messengers, explicit models can be constructed. Some examples of  spectra are given in section \ref{MODELBUILDING}. We have taken very close  or equal values for $F$ and $D$ terms, and chosen specific forms for the messenger couplings that forbid the presence of tachyonic adjoint scalars. The latter turn out to be the heaviest states of the models. For two of our examples they are out of LHC reach, and the DG-MSSM would appear as the MSSM in disguise. The third example has scalar gluon partners (sgluon) that can be produced at the LHC. In all cases, the $F$ terms used, with the assumption of minimal supergravity couplings would imply a gravitino mass of order 1 GeV.  As we are working in a global supersymmetric limit, we do not have explicit assumptions for the K\"ahler potential, or other $F$ terms, but these could lead to sequestered supergravity effects and might give a heavier mass for the gravitino.

\section{The Model at Low Energies}
\label{LOWENERGY}

\begin{table}[!ht]
\begin{center}
\begin{tabular}{c|c|c|c|c|c}
\hline
\small{Names}  &                 & Spin 0                  & Spin 1/2 & Spin 1 & \scriptsize{$(SU(3), SU(2), U(1)_Y)$} \\
\hline
&  &   &      &  &   \\
\small{Quarks}  & $\mathbf{Q}$   & $\tilde{Q}=(\tilde{u}_L,\tilde{d}_L)$  & $(u_L,d_L)$ & & (\textbf{3}, \textbf{2}, 1/6) \\
& $\mathbf{u^c}$ & $\tilde{u}^c_L$              & $u^c_L$     & & ($\overline{\textbf{3}}$, \textbf{1}, -2/3) \\
\small{($\times 3$ families)} & $\mathbf{d^c}$ & $\tilde{d}^c_L$     & $u^c_L$     & & ($\overline{\textbf{3}}$, \textbf{1}, 1/3)  \\
\hline
\small{Leptons} & $\mathbf{L}$ & ($\tilde{\nu}_{eL}$,$\tilde{e}_L$) & $(\nu_{eL},e_L)$ & & (\textbf{1}, \textbf{2}, -1/2) \\
\small{($\times 3$ families)} & $\mathbf{e^c}$ & $\tilde{e}^c_L$    & $e^c_L$          & & (\textbf{1}, \textbf{1}, 1)  \\
\hline
\small{Higgs} & $\mathbf{H_u}$ & $(H_u^+ , H_u^0)$ & $(\tilde{H}_u^+ , \tilde{H}_u^0)$ & & (\textbf{1}, \textbf{2}, 1/2)  \\
& $\mathbf{H_d}$ & $(H_d^0 , H_d^-)$ & $(\tilde{H}_d^0 , \tilde{H}_d^-)$ & & (\textbf{1}, \textbf{2}, -1/2) \\
\hline
\small{Gluons} & $\mathbf{W_{3\alpha}}$ & & $\lambda_{3\alpha} $                       & $g$              & (\textbf{8}, \textbf{1}, 0) \\
&   & & $  [\equiv \tilde{g}_{\alpha}]$                       &                &  \\
&  &   &      &  &   \\
W    & $\mathbf{W_{2\alpha}}$ & & $\lambda_{2\alpha} $ & $W^{\pm} , W^0$  & (\textbf{1}, \textbf{3}, 0) \\
&   & &  $ [\equiv \tilde{W}^{\pm} , \tilde{W}^{0}]$ &   &   \\
&  &   &      &  &   \\
B    & $\mathbf{W_{1\alpha}}$ & & $\lambda_{1\alpha} $                       & $B$              & (\textbf{1}, \textbf{1}, 0 ) \\
&   & & $ [\equiv \tilde{B}]$                       &                &    \\
\hline
\hline
\small{DG-octet}& $\mathbf{O_g}$ &  $O_g $  & $\chi_{g} $ &  & (\textbf{8}, \textbf{1}, 0) \\
&  $ [\equiv \mathbf{\Sigma_3}]$ &  $ [\equiv \Sigma_3]$  &  $  [ \equiv \tilde{g}']$ &  &  \\
&  &   &      &  &   \\
\small{DG-triplet} & $\mathbf{T}$ & $\{T^0, T^{\pm}\}$ & $\{\chi_T^0, \chi_T^{\pm}\}$ &  & (\textbf{1},\textbf{3}, 0 )\\
& $ [\equiv \mathbf{\Sigma_2}]$   &  $[ \equiv \Sigma_2]$ & $[ \equiv \{\tilde{W}'^{\pm},\tilde{W}'^{0}\}]$ &  &  \\
&  &   &      &  &   \\
\small{DG-singlet}  & $\mathbf{S}$& $S$ & $\chi_{S} $   &  & (\textbf{1}, \textbf{1}, 0 ) \\
&  $ [\equiv \mathbf{\Sigma_1}]$ & $ [\equiv \Sigma_1]$ &  $ [ \equiv \tilde{B}']$    &  &   \\
\hline
\hline
\small{Mass(GeV)}  &   &  & & &  \\
\hline
$10^{13}$ & $\mathbf{Q_{i3}}$ ($\times  2$)  &  & &  &  (\textbf{3}, \textbf{1}, 1/2) \\
&  $ \mathbf{\tilde{Q}_{\ov{i3}}}$($\times  2$)&   & & & ($\overline{\textbf{3}}$, \textbf{1}, -1/2) \\
\hline
$1.3\  10^{13} $ & $\mathbf{Q_{i2}}$ ($\times  4$ )     &  & & &  (\textbf{1}, \textbf{2}, 1/2) \\
& $ \mathbf{\tilde{Q}_{\ov{i2}}}$  ($\times  4$ ) & & & &  ($\overline{\textbf{1}}$, \textbf{2}, -1/2) \\
\hline
$3 \ 10^{12} $  & $\mathbf{Q_{i1}}$ ($\times 4$ )   &   & & & (\textbf{1}, \textbf{1}, 1) \\
& $ \mathbf{\tilde{Q}_{\ov{i1}}}$($\times 4$ ) &  &  & & ($\overline{\textbf{1}}$, \textbf{1}, -1) \\
\hline
\hline
\hline
&  &                & &     & Auxiliary  \\
\hline
U(1)'   &$\mathbf{W'_{\alpha}}$ &  &   & & D \\
\hline
\small{Spurion } & $\mathbf{X}$ &  & & & F \\
\hline
\end{tabular} 
\caption{Chiral and gauge multiplet fields in the model.}
\label{diracgauginos_Fields}
\end{center}
\end{table}

The  particle content of the model is presented in table \ref{diracgauginos_Fields} where one sees that in order to allow Dirac masses for the gauginos, the usual MSSM content is  extended by states in the adjoint representations, the ``DG-adjoints'' :
\begin{eqnarray}
\mathbf{S} & = & S + \sqrt{2} \theta \chi_S + \cdots  \\
\mathbf{T} & = & T  + \sqrt{2} \theta \chi_T + \cdots \\
\mathbf{O_g} & = & {O}_g  + \sqrt{2} \theta \chi_g + \cdots
\end{eqnarray}
where $S = \frac{1}{\sqrt{2}}(S_P+i S_M)$ is a singlet and $T= \sum_{a=1,2,3} T^{(a)}$ an $SU(2)$ triplet.  We also listed the messenger states and the spurions as source of supersymmetry breaking. In addition to the content in the table there is a ``hidden sector'' responsible for the generation of  the corresponding $D$, $F$  or both terms in order to break supersymmetry. For the purpose of this paper we assume that it is possible to parametrize their effects just through the spurion couplings. Note also the absence of right-handed neutrinos in table \ref{diracgauginos_Fields}; we shall not discuss neutrino masses in this work.

After integrating out the messengers, the low energy effective Lagrangian can be written as:
\begin{equation}
\mathcal{L}_{observable}= \mathcal{L}_{SUSY}+\C{L}_{Breaking}
\label{Lagrangian}
\end{equation}
The model  has soft terms that are not usually considered in the MSSM. In fact, the supersymmetry breaking part can  be split as  
\beq
\C{L}_{Breaking} =\C{L}_{Breaking}^{Standard}+\C{L}_{Breaking}^{\mathrm{Non-standard}}
\eeq
where the standard part contains terms of the type:
\beq
-\C{L}_{Breaking}^{Standard} = (m^2)^j_i \phi^i \phi_j + (\frac{1}{6} a^{ijk} \phi_i \phi_j \phi_k + \frac{1}{2} b^{ij} \phi_i \phi_j + \frac{1}{2} M_a \lambda_a \lambda_a + h.c. )
\eeq
Here lowered indices are components from chiral superfields, and $\phi^i = \phi_i^\dagger$. But there are additional non-standard terms that may also be soft:
\beq
-\C{L}_{Breaking}^{\mathrm{Non-standard}} = t^i \phi_i + \frac{1}{2} r_i^{jk} \phi^i \phi_j \phi_k  + m_D^{ia} \psi_i \lambda_a  + h.c.
\eeq
We shall discuss below all these terms. We will always assume $R$-parity conservation.

\subsection{The supersymmetric terms}
The supersymmetric part of the Lagrangian can be split into four parts:
\beq
\C{L}_{observable}= \C{L}_{gauge} + \bigg(\int d^2 \theta \, [ W_{Yukawa}+W_{Higgs} + W_{Adjoint}] + h.c. \bigg).
\label{Lagrangian2}
\eeq
The first contains the gauge interactions:
\begin{eqnarray}
\mathcal{L}_{gauge}= & &\int  d^2\theta  \left[ \right.   \frac{1}{4}  \mathbf{W}_{1}^{\alpha}  \mathbf{W}_{1\alpha} + \frac{1}{2} \textrm{tr}(\mathbf{W}_{2}^{\alpha}  \mathbf{W}_{2\alpha}) + \frac{1}{2} \textrm{tr}(\mathbf{W}_{3}^{\alpha}  \mathbf{W}_{3\alpha})] + h.c. 
\nonumber \\ 
&+&\int  d^2\theta d^2{\bar{\theta}}\quad (\sum_{ij}\mathbf{ \Phi}_{i}^\dagger e^{g_j \mathbf{V_j}} \mathbf{\Phi}_{i} + h.c.)
\end{eqnarray}
where $\mathbf{V}_j$ are the vector  and $\mathbf{W}_{j \alpha} $ the corresponding field strength superfields associated to $U(1)_Y$, $ SU(2)$ and $ SU(3)$  for $j=1, 2, 3$ respectively.

The  second part contains the Yukawa  superpotential for the MSSM matter fields:
\begin{eqnarray}
W_{Yukawa}&=& Y^{ij}_U \mathbf{ Q_i \cdot H_u u^c_j} + Y^{ij}_D \mathbf{ Q_i \cdot H_d d^c_j}
+  Y^{ij}_E \mathbf{L_i \cdot H_d e^c_j }  \nonumber 
\label{Yukawa_Superpotential}
\end{eqnarray}
where the bold characters denote superfields. Here, $i,j$ are family indices and run from $1$ to $3$. The $3 \times 3$ $Y$ matrices are the Yukawa couplings. The ``$\cdot$'' denotes 
$SU(2)$ invariant couplings, for example: $Q \cdot H_u = \tilde{u}_L H_u^0 - \tilde{d}_L H_u^+ $.

The DG-adjoints  modify  the Higgs  superpotential, since new relevant and marginal operators are now allowed:
\begin{align}
W_{Higgs}=& \mu \mathbf{H_u\!\cdot\! H_d }   + \lambda_S \mathbf{SH_d\!\cdot\! H_u}  + 2  \lambda_T \mathbf{H_d\!\cdot\! T H_u} 
\label{NewSuperPotential}
\end{align}
with the definition $H_u\cdot H_d = H^+_uH^-_d - H^0_u H^0_d$. Note that if there is an $N = 2$ extension of the gauge sector at some scale (such as the GUT scale), and if the Higgs multiplets $H_u$ and $H_d$ form an $N = 2$ hypermultiplet then  $\lambda_S$ and $ \lambda_T$ are related to the gauge couplings, at the $N=2$ scale, by:
\begin{equation}
\lambda_S= \sqrt{2} g_Y \frac{1}{2} , \qquad  \lambda_T= \sqrt{2} g_2 \frac{1}{2} , 
\label{Neq2Lambdas}\end{equation}
where $g_Y$ and $g_2$ are the  $U(1)_Y$ and $SU(2)$ gauge couplings respectively. The factor $1/2$ in $\lambda_S$ arises from the $U(1)_Y$ charge of the Higgs doublets.

Finally, the most general renormalisable superpotential involving only adjoint fields is
\begin{align}
W_{Adjoint} =& L \mathbf{S} + \frac {M_S}{2}\mathbf{S}^2 + \frac{\kappa_S}{3} \mathbf{S}^3 + M_T \textrm{tr}(\mathbf{TT}) +\lambda_{ST} \mathbf{S}\textrm{tr}(\mathbf{TT}) \nonumber \\
& + M_O \textrm{tr}(\mathbf{OO})+\lambda_{SO} \mathbf{S}\textrm{tr}(\mathbf{OO})  + \frac{\kappa_O}{3} \textrm{tr}(\mathbf{OOO}).
\label{AdjointSuperpotential}\end{align}
Note there are no terms $\textrm{tr}(\mathbf{T}),\textrm{tr}(\mathbf{O}), \textrm{tr}(\mathbf{TTT})$ since these vanish by gauge invariance. The dimensionful quantities $L, M_S, M_T$ and $M_O$, in particular if they are assumed to take values of order of the electroweak scale, potentially introduce an issue of scale hierarchy in the same way as the Higgs $\mu$-term. As we shall briefly discuss in section \ref{CONSTRAINTS:NoMajoranas}, the supersymmetric masses could be generated by loops. 

We shall generally be assuming (or, in the case of the masses, ensuring) that all of the terms in $W_{Adjoint}$ vanish; this is because in order to preserve R-symmetry with Dirac gaugino masses the adjoint superfields must have R-charge zero. Alternatively we could assume an underlying $N=2$ supersymmetry at the GUT scale. However, it is an interesting possibility (though not one that we shall exploit) that the only source of R-symmetry breaking in the model is an explicit small dimensionless coupling (such as $\kappa_S$) which would allow the supersymmetry breaking sector to preserve an R-symmetry whilst allowing for the absence of an R-axion.

\subsection{The standard soft terms}

The supersymmetry breaking soft terms denoted as "standard" can be separated into four parts : (i) soft terms involving only the MSSM matter fields (ii) standard terms for the DG-adjoint fields (iii) possible A-terms and (iv) ($R$-symmetry breaking) Majorana gaugino masses:
\begin{eqnarray}
-\C{L}_{Breaking}^{Standard} &= & -\mathcal{L}^{MSSM}_{soft} -\Delta\mathcal{L}_{soft}^{DG-Adjoint}- \Delta \mathcal{L}_{soft}^{A} -\C{L}_{Gaugino}^{Majorana}
\end{eqnarray}
where
\begin{eqnarray}
-\mathcal{L}^{MSSM}_{soft} &= &   \tilde{Q}^\dagger_i  {m_Q^2}_{ij} \tilde{Q}_j+ \tilde{L}^\dagger_i  {m_L^2}_{ij} \tilde{L}_j+
\tilde{u}_i  {m_u^2}_{ij} \tilde{u}_j^\dagger + \tilde{d}_i  {m_d^2}_{ij} \tilde{d}_j^\dagger + \tilde{e}_i  {m_e^2}_{ij} \tilde{e}_j^\dagger\nonumber \\ && + A_U^{ij} { \tilde{Q}_i \cdot H_u \tilde{u}^c_j} + A_D^{ij} {\tilde{Q}_i \cdot H_d \tilde{d}^c_j }
+ A_E^{ij} {\tilde{L}_i \cdot H_d  \tilde{e}^c_j }\nonumber  \\ && + m_{H_u}^2 |H_u|^2 + m_{H_d}^2 |H_d|^2 + B_\mu (H_u\cdot H_d +c.c.)
\end{eqnarray}
and
\begin{eqnarray}
- \Delta\mathcal{L}_{soft}^{DG-Adjoint} &= &  m_S^2  |S|^2 + \frac{1}{2} B_S (S^2 + h.c.)  + 2 m_T^2 \textrm{tr}(T^\dagger T) + B_T (\textrm{tr}(T T)+ h.c.) \nonumber \\ &&+ 2 m_O^2 \textrm{tr}(O^\dagger O) + B_O (\textrm{tr}(OO)+ h.c.)  . 
\label{Lsoft-DGAdjoint}
\end{eqnarray}
The adjoint scalar A-terms (including the possible scalar tadpole) are given by
\begin{align}
- \Delta \mathcal{L}_{soft}^{A} =& A_S  \lambda_S SH_d\cdot H_u +  2 A_T  \lambda_T H_d \cdot T H_u + t^S S + \frac{1}{3} \kappa_S  A_{\kappa_S} S^3 \nonumber \\
& + \lambda_{ST} A_{ST} S \tr(TT) + \lambda_{SO} A_{SO} S \tr(OO) + \frac{1}{3} \kappa_O A_{\kappa_O} \tr(OOO) \nonumber \\
&+ h.c.   
\end{align}
However, to allow Dirac gauginos the only global symmetry that the adjoint field may transform under is $R$-symmetry, since the gaugino must also tranform; if it is preserved by the adjoint fields this excludes the $A_S, A_T$ terms, and also, since it requires $\lambda_{ST} = \lambda_{SO} = \kappa_O = 0$ we will generically not generate $A_{ST}, A_{SO}, A_{\kappa_O}$ terms.  
Finally
\begin{align}
\mathcal{L}_{Gaugino}^{Majorana}=&\int  d^2\theta  \left[ \right.   \frac{1}{4} \textbf{M}_{1} \mathbf{W}_{1}^{\alpha}  \mathbf{W}_{1\alpha} + \frac{1}{2}\textbf{M}_{2} \textrm{tr}(\mathbf{W}_{2}^{\alpha}  \mathbf{W}_{2\alpha}) + \frac{1}{2}\textbf{M}_{3} \textrm{tr}(\mathbf{W}_{3}^{\alpha}  \mathbf{W}_{3\alpha})  \left.  \right]
\label{Majoranagauge}
\end{align}
where  we have introduced  spurion superfields to take into account the possibility of Majorana gaugino masses:
\begin{eqnarray}
\textbf{M}_{i} &= &  2 \theta \theta M_{i}
\end{eqnarray}
which require $R$-symmetry breaking. Unless stated otherwise, we will take 
\begin{eqnarray}
\textbf{M}_{i} &= &  0
\end{eqnarray}

\subsection{The non-standard soft terms}

The possible non-standard soft terms are
\begin{align}
-\C{L}^{\mathrm{Non-standard}} =& m_{1D} \chi_S \lambda_Y + 2 m_{2D} \tr(\chi_T \lambda_2)+ 2 m_{3D} \tr(\chi_O \lambda_3) \nonumber \\
&+ \tilde{Q}^\dagger_i r^{S\tilde{Q}_j}_{\tilde{Q}_i}  S \tilde{Q}_j+  \tilde{L}^\dagger_i  r^{S\tilde{L}_j}_{\tilde{L}_i} S \tilde{L}_j+
\tilde{u}_j r^{S\tilde{u}_j}_{\tilde{u}_i} S \tilde{u}_i^\dagger +  \tilde{d}_j r^{S\tilde{d}_j}_{\tilde{d}_i}S \tilde{d}^\dagger_i+  \tilde{e}_j r^{S\tilde{e}_j}_{\tilde{e}_i} S\tilde{e}^\dagger_i\nonumber \\
& +\tilde{H_u}^\dagger  r^{S\tilde{H_u}}_{\tilde{H_u}}S \tilde{H_u}+\tilde{H_d}^\dagger  r^{S\tilde{H_d}}_{\tilde{H_d}}S \tilde{H_d} \nonumber \\
&+(\tilde{Q}^\dagger_i r^{T_a\tilde{Q}_j}_{\tilde{Q}_i} T_a \tilde{Q}_j) + (\tilde{L}^\dagger_i r^{T_a\tilde{L}_j}_{\tilde{L}_i} T_a  \tilde{L}_j) +   (\tilde{H_u}^\dagger r^{T_a\tilde{H_u}}_{\tilde{H_u}}T_a \tilde{H_u}) + (\tilde{H_d}^\dagger r^{T_a\tilde{H_d}}_{\tilde{H_d}}T_a \tilde{H_d})
\nonumber \\
&+ (\tilde{Q}^\dagger_i r^{O_a\tilde{Q}_j}_{\tilde{Q}_i} O_a  \tilde{Q}_j)+  (\tilde{u}_j r^{O_a\tilde{u}_j}_{\tilde{u}_i}O_a \tilde{u}^\dagger_i)+ ( \tilde{d}_j r^{O_a\tilde{d}_j}_{\tilde{d}_i}O_a \tilde{d}^\dagger_i )\nonumber \\
&+ r^{SS}_S S^\dagger S^2 + r^{T^aT^b}_S S^\dagger \tr(T^aT^b) + r^{ST^a}_{T^b} S \tr((T^b)^\dagger T^a) + r^{T^a T^b}_{T^c} \tr(T^a T^b (T^c)^\dagger)\nonumber \\
&+ r^{O^aO^b}_{O^c} \tr((O^c)^\dagger O^aO^b) + r^{O^aO^b}_S S^\dagger \tr(O^aO^b) + r^{SO^a}_{O^b} S\tr((O^b)^\dagger O^a) \nonumber \\
& + h.c.
\end{align}
This is a large number of new terms, but it transpires that in a model of Dirac gauginos with spontaneously broken supersymmetry there are relations amongst them, as we shall discuss.

The Dirac gaugino masses  arise from the Lagrangian:
\begin{align}
\mathcal{L}^{Dirac}_{gaugino}=& \int  d^2\theta  \left[ \right.    \sqrt{2} \textbf{m}^\alpha_{1D} \mathbf{W}_{1\alpha} \mathbf{S}  
+  2\sqrt{2} \textbf{m}^\alpha_{2D} \textrm{tr}(\mathbf{W}_{2 \alpha} \mathbf{T}) +  2\sqrt{2} \textbf{m}^\alpha_{3D} \textrm{tr}(\mathbf{W}_{3 \alpha} \mathbf{O_g}) \left.  \right] \nonumber \\
& +h.c. 
\label{Newdiracgauge}
\end{align}
where we have introduced  spurion superfields to parametrize the generation of Dirac gaugino masses:
\begin{eqnarray}
\textbf{m}_{\alpha iD} &= & \theta_\alpha m_{iD}.
\end{eqnarray}

The non-standard soft terms must then all arise from  the holomorhic term (\ref{Newdiracgauge}). Integration on the spinor coordinates, and going on-shell, leads to  Dirac masses as well as new interactions
\beq
\int d^2 \theta \sqrt{2} m_D \theta^\alpha W_\alpha \Sigma \supset - m_D (\lambda \psi) + \sqrt{2} m_D \Sigma D
\eeq
for $U(1)$ gauginos (with $D$ being the $D$-term of the gauge group), or
\beq
\int d^2 \theta 2\sqrt{2} m_D \theta^\alpha \tr (W_\alpha \Sigma) \supset - m_D (\lambda_a \psi_a) + \sqrt{2} m_D \Sigma_a D_a
\eeq
for $SU(N)$. Then with $D^a_b = - g_b \phi^\dagger_i R^a_b (i) \phi_i $ (where $R_b^a (i) $ is the $a^{th}$ generator of the group $b$ in the representation of field $i$, and $R_Y^b (i) = Y (i)$ for the hypercharge) we find
\beq
\C{L} \supset - m_{bD} \sqrt{2}  g_b \Sigma_a \phi^\dagger R^a_b \phi
\eeq
and thus for fields in the fundamental or antifundamental representations
\beq
r_i^{i\Sigma_a} = m_{bD} \sqrt{2}  g_b R^a_b (i) .
\label{req}
\eeq
Since these couplings come from a holomorphic term, this relation is preserved by the renormalisation group running, which has been confirmed up to two loops \cite{Jack:1999fa}. We also find that there are no couplings of the form $r^{SS}_S, r^{TT}_T$ or $r^{OO}_O$ because 
\beq
D_a \supset - i f^{abc} \Sigma^b (\Sigma^\dagger)^c  \rightarrow \C{L} \supset  - i m_D \sqrt{2} g \Sigma_a  \Sigma_b (\Sigma^\dagger)_c f^{abc} = 0,
\eeq
and clearly no mixed terms $r^{ST}_T, r^{SO}_O, r^{TT}_S, r^{OO}_S$ are generated.

Note also that there is a supersymmetric term that mimics $r^{jk}_i \,:\, y^{jkl} \mu_{il} $; this appears in the terms for the Higgs.  For instance, from the $F$-term  of $H_u$,  $|\mu H_d + \lambda_S SH_d|^2$ which gives rise to $\lambda_S \mu^* H_d^\dagger S H_d$.

Nonstandard soft supersymmetry breaking terms are often neglected and considered to be ``hard'' because they may generate quadratic divergences in tadpole terms. Of course, if there are no gauge singlets in the model then these are absent, but since we are considering Dirac gaugino masses then the $U(1)$ adjoint is such a singlet. However, if we break supersymmetry spontaneously, then the sum of all contributions to the quadratic divergences cancels; we can easily see this at one loop, since the quadratic divergence is proportional to $\sum_i r^{Si}_i$, and provided that anomalies are cancelled, $\sum_i r^{Si}_i = m_D \sqrt{2}  g_Y \sum_i Y_i =0$.

\subsection{Higgs Sector and R Symmetry Breaking}
\label{Higgs}

Since our low energy theory contains adjoint fields with couplings to the Higgs sector, we must examine how these affect the electroweak symmetry breaking. In fact, there exist several possibilities for Higgs phenomenology differing from the MSSM, such as
\begin{itemize}
\item $\cancel{\mu}$MSSM (MSSM without $\mu$ term) \cite{Nelson:2002ca}. This requires a large $\lambda_T (\lesssim 0.6)$ and low Dirac Wino mass $\sim 100$ GeV to allow sufficiently heavy charginos. $R$ symmetry is explicitly broken by a small $B_\mu$ term, possibly arising from gravitational interactions. 
\item Along the lines of the NMSSM, requiring additional adjoint superpotential terms and tachyonic soft mass for the singlet scalar $S$. Since in gauge mediation the adjoint scalars are typically the heaviest states, of particular interest to us would be the case of a heavy singlet with a small cubic superpotential coupling $W \supset \frac{\kappa_S}{3} S^3$. This term explicitly violates $R$ symmetry, and thus this could be the only source of R-breaking in the model. The singlet would obtain a large vev and a large physical mass of $-m_S^2$, and so we could integrate it out, leaving effective $\mu, B_\mu$-terms $\mu_{eff}^2 = \frac{-\lambda_S^2 m_S^2}{2\kappa^2_S}, B_{\mu eff} = \frac{-m_S^2 \lambda_S}{2\kappa_S}$. We could then take $\lambda_S \sim \kappa_S$ and both couplings small.
\end{itemize}
However, we shall leave the realisation of these scenarios to future work, and instead take a more conservative approach: as in \cite{Belanger:2009wf}, we shall consider models with all adjoint scalar soft masses  large and real, so that they can be integrated out;  following \cite{Nelson:2002ca} we shall suppose that higher-dimension operators generate a small $B_\mu$ term, but we differ by also adding a $\mu$ term as necessary for electroweak symmetry breaking, as dictated by the soft Higgs masses. We then obtain the MSSM supplemented by Dirac gaugino masses and a modified Higgs potential.

Using the standard form of the Higgs potential
\begin{eqnarray}
\label{potential4}
V_{eff} & = & (m_{H_u}^2+\mu^2) |H_u|^2 +
(m_{H_d}^2+\mu^2) |H_d|^2
         - [m_{12}^2 H_u\cdot H_d + h.c. ] \nonumber \\
   &   & + \frac{1}{2}\big[\frac{1}{4}(g^2+g'^2) + \lambda_1\big]
                       (|H_d|^2)^2
         +\frac{1}{2}\big[\frac{1}{4}(g^2+g'^2) + \lambda_2\big]
                       (|H_u|^2)^2 \nonumber \\
   &  &  +\big[\frac{1}{4}(g^2-g'^2) + \lambda_3\big]
                       |H_d|^2 |H_u|^2
         +\big[-\frac{1}{2}g^2 + \lambda_4\big]
(H_d\cdot H_u)(H_d^*\cdot H_u^*)\nonumber\\
   & &   +\big(\frac{\lambda_5}{2} ( H_d\cdot H_u)^2
              +\big[ \lambda_6 |H_d|^2 + \lambda_7 |H_u|^2\big]
                    ( H_d\cdot H_u) + h.c. \big),
\end{eqnarray}
we find $\lambda_3$ and $\lambda_4$ have a tree level contribution, so we write $\lambda_3 = 2\lambda_T^2 + \lambda_3^\prime, \lambda_4 = \lambda_S^2 - \lambda_T^2 + \lambda_4^\prime$ where $\lambda_3^\prime, \lambda_4^\prime$ are the loop corrections to the potential. Then assuming that $B_S, B_T$ are small, $\lambda_4^\prime =\lambda_5 = \lambda_6 = \lambda_7 = 0,$ and
\begin{eqnarray}
\lambda_1 &\approx& \frac{3}{16\pi^2} y_b^4 \log \left(\frac{m_{\tilde{b}_1}^2m_{\tilde{b}_2}^2 }{v^4} \right) + \frac{5}{16\pi^2} \lambda_T^4 \log \left(\frac{m_T^2}{v^2}\right) + \frac{1}{16\pi^2} \lambda_S^4 \log \left(\frac{m_S^2}{v^2}\right) \nonumber \\
&&- \frac{1}{16\pi^2} \frac{\lambda_S^2 \lambda_T^2}{m_T^2 - m_S^2} \bigg\{ m_T^2 [\log \left(\frac{m_T^2}{v^2}\right) -1] - m_S^2 [\log \left(\frac{m_S^2}{v^2}\right) -1] \bigg\}\nonumber \\
\lambda_2 &\approx& \frac{3}{16\pi^2} y_t^4 \log \left(\frac{m_{\tilde{t}_1}^2m_{\tilde{t}_2}^2 }{m_t^4} \right) +\frac{5}{16\pi^2} \lambda_T^4 \log \left(\frac{m_T^2}{v^2}\right) + \frac{1}{16\pi^2} \lambda_S^4 \log \left(\frac{m_S^2}{v^2}\right) \nonumber \\
&&- \frac{1}{16\pi^2} \frac{\lambda_S^2 \lambda_T^2}{m_T^2 - m_S^2} \bigg\{ m_T^2 [\log \left(\frac{m_T^2}{v^2}\right) -1] - m_S^2 [\log \left(\frac{m_S^2}{v^2}\right) -1] \bigg\}\nonumber \\ 
\lambda_3^\prime &\approx& \frac{5}{32\pi^2} \lambda_T^4 \log \left(\frac{m_T^2}{v^2}\right) + \frac{1}{32\pi^2} \lambda_S^4 \log \left(\frac{m_S^2}{v^2}\right) \nonumber \\
&&+ \frac{1}{32\pi^2} \frac{\lambda_S^2 \lambda_T^2}{m_T^2 - m_S^2} \bigg\{ m_T^2 [\log \left(\frac{m_T^2}{v^2}\right) -1] - m_S^2 [\log \left(\frac{m_S^2}{v^2}\right) -1] \bigg\},
\end{eqnarray} 
where $y_t, y_b$ are the top and bottom Yukawa couplings in the third-family dominant approximation. This then gives us the minimisation conditions
\begin{align}
\mu^2 =& -\frac{M_Z^2}{2} + \frac{1}{\tan^2 \beta - 1} ( m_{H_d}^2 + \Delta_d - \tan^2 \beta (m_{H_u}^2 + \Delta_u) ) \nonumber \\
m_{\tilde{A}}^2 =& - \frac{2m_{12}^2}{\sin 2\beta} = m_{H_u}^2 +  m_{H_u}^2 + 2\mu^2 + \frac{2 (\lambda_S^2 + \lambda_T^2)}{g_Y^2+g_2^2} M_Z^2 + \Delta_u + \Delta_d > 0,
\end{align}
where for $v \simeq 246$ GeV, $\tan \beta = \bra H_u \ket/\bra H_d \ket$,
\begin{align}
\Delta_u \equiv&\frac{1}{2} \lambda_2 v^2 \sin^2 \beta +\frac{1}{2} \lambda_3^\prime v^2 \cos^2 \beta \nonumber \\
\Delta_d \equiv& \frac{1}{2} \lambda_1 v^2 \cos^2 \beta +\frac{1}{2} \lambda_3^\prime v^2 \sin^2 \beta.
\end{align}

We also have low energy corrections to the Higgs masses from integrating out the heavy squarks 
\begin{align}
\Delta m_{H_u}^2 \approx \frac{1}{16\pi^2} \bigg[ &3 |y_t|^2 m_{Q_3}^2 \log \frac{m_{Q_3}^2}{m_t^2} + 3 |y_t|^2 m_{U_3}^2 \log \frac{m_{U_3}^2}{m_t^2} \nonumber \\
&+ \lambda_S^2 m_S^2 \log \frac{m_S^2}{v^2} + 3\lambda_T^2 m_T^2 \log \frac{m_T^2}{v^2} \bigg] \nonumber \\
\Delta m_{H_d}^2 \approx \frac{1}{16\pi^2} \bigg[&  3 |y_b|^2 m_{Q_3}^2 \log \frac{m_{Q_3}^2}{v^2} +   3 |y_b|^2 m_{D_3}^2 \log \frac{m_{D_3}^2}{v^2} +     |y_\tau|^2 m_{L_3}^2 \log \frac{m_{L_3}^2}{v^2} \nonumber \\
&+    |y_\tau|^2 m_{E_3}^2 \log \frac{m_{E_3}^2}{v^2} + \lambda_S^2 m_S^2 \log \frac{m_S^2}{v^2} + 3\lambda_T^2 m_T^2 \log \frac{m_T^2}{v^2} \bigg].
\end{align}

\subsubsection{T Parameter}

As computed in \cite{Belanger:2009wf}, in a model where soft adjoint scalar masses are significantly above the electroweak scale, the tree-level correction to the electroweak precision variable $T$ is  
\begin{align}
\Delta \rho \simeq&  \bigg[\frac{v}{(M_T^2+m_T^2+4 m^2_{2D}+ B_T)} \bigg]^2 \nonumber \\
&\times \left[ - g    m_{2D}  \cos 2\beta -{\sqrt{2}} \mu    \lambda_T + \frac{\lambda_T }{\sqrt{2}} (  M_T+    A_T)  \sin 2\beta  \right] \bigg]^2,
\end{align}
where $\rho = 1 + \alpha T = 1.0004^{+0.0008}_{-0.0004}$~\cite{Amsler:2008zzb} and $v \simeq 246 \, \mathrm{GeV}$. For our model  $M_T = A_T = 0$.

\section{Gauge Mediated Soft Masses}
\label{SOFTMASSES}

In this section we shall disuss the generation of soft supersymmetry breaking masses for the low energy theory in the previous section via gauge mediation. We shall assume vector-like pairs of messengers $N_a \times (\mathbf{Q_{ia}}, \mathbf{ \tilde{Q}_{\ov{j}a}})$ in the fundamental and antifundamental representations of group $a$ with degenerate masses, and with hypercharges $Y_a, -Y_a$. The couplings of the messengers to the adjoints $\Sigma_a$ and the $F$-term spurion $\mathbf{X}$ (having vev $\bra \mathbf{X} \ket = \theta^2 F$) are
\begin{align}
\C{L}^{Mess}_{F}  =&\int d^2\theta   [\Mm^{(a)} \mathrm{tr} (\mathbf{Q_{ia}}\mathbf{ \tilde{Q}_{\ov{j}a}}) \delta_{i\ov{j}} + \lambda^{(ab)}_{i\ov{j}} \mathrm{tr}(\mathbf{Q_{ia} \Sigma_b \tilde{Q}_{\ov{j}a}}) + \kappa^{(a)}_{i\ov{j}} \mathrm{tr}(\mathbf{Q_{ia}\tilde{Q}_{\ov{j}a}}) \mathbf{X}].
\label{FtermW}
\end{align}
Here $\lambda^{(ab)}_{i\ov{j}}, \kappa^{(a)}_{i\ov{j}}$ are matrices  with messenger family indices, that will be constrained in the later sections. We shall assume that the choice of couplings is such that R-symmetry is preserved (thus for example the F-term spurion $\mathbf{X}$ must have R-charge $2$ and so cannot couple diagonally to the messengers, and all terms $\kappa^{(a)}_{i\ov{i}} $ vanish). 

The (only) partial degeneracy of the masses will be explained in later sections. We shall consider first the contributions to  soft masses by $D$ terms, and then by $F$-terms. We shall take as motivation the preservation of R-symmetry by the supersymmetry breaking sector, and so shall not generate $A$ or $B_\mu$ terms; rather we shall assume that R-symmetry is broken either by explicit operators from a higher-energy theory, or explicit couplings in the visible sector, as  discussed in section \ref{Higgs}.

The messengers are assumed to carry charges $e^{(a)}_{i}$ under a hidden $U(1)'$ with potentially non-vanishing $D$-term, which couples to the messenger scalars as:
\begin{eqnarray}
\C{L}^{Mess}_{D}&=& D[\sum_{i,a}e^{(a)}_{i} \tr(Q_{ia} {Q}^\dagger_{{i}a} - \tilde{Q}_{ia} \tilde{Q}^\dagger_{{i}a})  ] \nonumber \\
&\equiv& D[  \sum_{a} \tr(Q_{ia} \hat{e}^{(a)}_{i\ov{j}}{Q}^\dagger_{{j}a} - \tilde{Q}_{ia} \hat{e}^{(a)}_{i\ov{j}}\tilde{Q}^\dagger_{{j}a})  ]
\label{LDp}
\end{eqnarray}
and leads to messenger mass splittings. Here we have written $D$ to denote the hidden $D$-term, but note that this absorbs a factor of the hidden gauge coupling; this is an unknown quantity, but we should bear in mind that the true supersymmetry breaking scale is somewhat larger than this. Note also that we shall \emph{not} insist that the messenger couplings $ \lambda^{(ab)}_{i\ov{j}} $ respect the hidden $U(1)'$ symmetry; we shall suppose that it is massive and that couplings that violate it are generated by expectation values of other unspecified standard-model singlet fields.

\subsection{$D$-terms}

We first consider the case where the source of breaking is the $D$ of a hidden (massive) $U(1)$.
To allow Dirac gaugino masses it is not only permitted but required that there are direct superpotential couplings between the messengers and the adjoints. The Dirac gaugino masses and the  adjoint masses are then both generated at one-loop by messengers that have soft supersymmetry breaking  as shifts in the scalar masses by the  $D$-terms given by $\C{L}^{Mess}_{D}$.
\beq
m_{bD} =  \frac{1}{\sqrt{2}}g_b \tr (\lambda^{(ab)} \hat{e}^{(a)} \hat{Y})  \sum_{i}\C{I} (\Mm^{(a)}, D) 
\label{DTermG}\eeq
where we have defined
\beq\begin{split}
\C{I} &\equiv \frac{2}{(4\pi)^2} \frac{D}{\Mm^{(a)}}  \bigg[ \frac{(1-\frac{D}{(\Mm^{(a)})^2}) \log (1-\frac{D}{(\Mm^{(a)})^2}) + (1+\frac{D}{(\Mm^{(a)})^2})\log (1+\frac{D}{(\Mm^{(a)})^2})}{D^2/(\Mm^{(a)})^4} \bigg] \\
&= \frac{2}{(4\pi)^2} \frac{D}{\Mm^{(a)}} \bigg[ 1 + \C{O}\left(\frac{D^2}{(\Mm^{(a)})^4}\right) \bigg].
\end{split}\label{DefI}\eeq

The masses for the adjoint scalars $\Sigma$ are given by \cite{Benakli:2008pg}
\begin{align}
-\mathcal{L} \supset& m_\Sigma^2  2^\delta \tr(\Sigma^\dagger \Sigma) + \frac{1}{2} B_\Sigma 2^\delta \tr(\Sigma^2 + (\Sigma^\dagger)^2) \nonumber\\
m_\Sigma^2 =& 2^{-\delta} \frac{1}{96\pi^2} \frac{D^2}{\Mm^2}  \mathrm{tr} \bigg(  [\hat{e}, \lambda]([\hat{e},\lambda])^\dagger\bigg)+ 2^{-\delta} \frac{3D}{64\pi^2} \mathrm{tr} ( \hat{e} [\lambda,\lambda^\dagger])\nonumber \\
B_\Sigma =& -2 \times 2^{-\delta} \frac{1}{96\pi^2} \frac{D^2}{\Mm^2} \mathrm{tr} \bigg( 2 \lambda^2 \hat{e}^2 + \lambda \hat{e} \lambda \hat{e} \bigg)
\label{AdjointDtermMasses}\end{align}
where $\delta$ is $1$ for $SU(N)$ and $0$ for $U(1)$.

The real and imaginary components $\Sigma$ comprise the propagating degrees of freedom; writing $\Sigma \equiv \frac{1}{\sqrt{2}} (\Sigma_P + i \Sigma_M)$, we have 
\beq
-\mathcal{L} \supset 2^\delta \tr\bigg( \frac{1}{2} (m^2 + B) \Sigma_P^2 + \frac{1}{2} (m^2 - B) \Sigma_M^2 \bigg),
\label{Sadjoints}\eeq
and thus the physical masses are $m_{\Sigma_P}^2, m_{\Sigma_M}^2 = m_\Sigma^2 \pm B_\Sigma$.

The lowest order in $D/\Mm^2$ sfermion masses are generated at three loops 
\beq
m_{\tilde{f}}^2 = \sum_{b=1}^3 C_{\tilde{f}}^b \frac{ (m_{bD})^2 \alpha_b}{\pi} \log \left(\frac{m_{\Sigma_P}^{(b)}}{m_{bD}}\right)^2 ,
\label{Sfermions}
\eeq
where $C_{\tilde{f}}^b$ is the quadratic Casimir of the field $f$ under group $b$, equal to $Y^2(f)$ for $U(1)_Y$, and $\frac{N^2-1}{2N}$ for $SU(N)$. The above represents integrating out the adjoint fields, and so should be performed at low energies rather than the messenger scale. In a pure D-term breaking scenario, this leaves the sfermions to be the lightest states, providing the next to lightest supersymmetric particle (NLSP), the lightest being the (Dirac) gravitino. Model building in this way is then constrained by the low mass for the selectron due to the weakness of the hypercharge at low energies.

\subsection{$F$-terms}

The Dirac gaugino masses are generated at one-loop and can be expressed to leading order in $F$ as:
\beq
m_{bD} = \frac{g_b}{\sqrt{2}} \frac{1}{16\pi^2}\frac{|F|^2}{6(M_{\mathrm{Mess}}^{(a)})^3}  \mathrm{tr} ( I^b_{Q_{ia}} \lambda [ \kappa, \kappa^\dagger] ),
\label{Dirac_FTerm}\eeq
where $I^b_{Q_{ia}}$ is the Dynkin index  of the messengers $Q_{ia}$ under group $b$, equal to $1/2$ for fundamental-antifundamental $SU(N)$ messengers and $Y^2$ for $U(1)$ pairs. 

The soft masses for this model are given at two loops by
\beq
m_{\tilde{f}}^2 = 2 \sum_{b=1}^3 C_{\tilde{f}}^b  \left(\frac{\alpha_b}{4\pi} \right)^2 (\Lambda_S^{(ab)})^2
\eeq
where  
\beq
(\Lambda_S^{(ab)})^2 =  \frac{|F|^2}{(M_{mess}^{(a)})^2} \mathrm{tr} ( 2 I^b_{Q_{ia}} \kappa \kappa^\dagger).
\eeq

The adjoint masses are given by \cite{Benakli:2008pg}
\begin{align}
m_\Sigma^2 =& 2^{-\delta}\frac{1}{16\pi^2} \frac{F^\dagger F}{\Mm^2}  \frac{1}{6}\tr \bigg( 2[\lambda, \lambda^\dagger][\kappa ,\kappa^\dagger] + [\lambda, \kappa]([\lambda,\kappa])^\dagger \bigg)  \nonumber \\
B_\Sigma =& -2 \times 2^{-\delta}\frac{1}{16\pi^2} \frac{F^\dagger F}{\Mm^2} \times \frac{1}{6} \tr \bigg( \kappa^\dagger (\kappa \lambda^2 + \lambda \kappa \lambda + \lambda^2 \kappa)\bigg)
\label{AdjointFtermMasses}\end{align}
where as before $\delta$ is $1$ for $SU(N)$ and $0$ for $U(1)$.

The chief drawback of a pure $R$-symmetric $F$-term is clear from an operator analysis or equation (\ref{Dirac_FTerm}); the lowest order Dirac masses are of second order in $F$, and therefore can only be acceptably large for a low messenger scale. In which case, the couplings will typically not unify in a perturbative regime. Hence to alleviate this problem, and the converse problem of small selectron masses for pure D-term breaking, for our model-building efforts we shall consider a combination of D- and F-terms.

\subsection{Summary}

To allow Dirac gaugino masses generated at the leading order in the supersymmetry breaking parameter, and sufficiently heavy selectrons, we shall consider a combination of $D$- and $F$-term breaking, with both $D$- and $F$-terms comparable. This can be easily realised, for example, in the context of semi-direct gauge mediation \cite{Seiberg:2008qj}. This will generate a spectrum with masses of generic order of magnitude
\begin{itemize}
\item Gaugino masses $\sim \frac{\lambda g}{16\pi^2} \frac{D}{\Mm}$
\item Sfermion masses $\sim  \frac{g^2}{16\pi^2} \frac{F}{\Mm}$
\item Adjoint scalar masses $\sim \frac{\lambda}{4\pi} \frac{D}{\Mm},\frac{\lambda}{4\pi} \frac{F}{\Mm} $.
\end{itemize}
Thus we expect the adjoint scalars to be the most massive states.

\section{Model building constraints}
\label{CONSTRAINTS}

\subsection{No leading order Majorana masses}
\label{CONSTRAINTS:NoMajoranas}

The first condition we impose is that the messenger superpotential does not break $R$-symmetry inducing then a Majorana gaugino mass at leading order. The latter is of the form:
\beq
m_{\lambda\lambda}^r = \frac{\alpha_r}{4\pi} \frac{F}{\Mm} \mathrm{tr} (\kappa).
\eeq
whose absence implies:
\beq 
\mathrm{tr} (\kappa)= 0
\eeq
This is independent of the adjoints, although if we have $R$-symmetry violating couplings in $\lambda$ we expect subleading Majorana masses at higher loop order. However, we expect them to contribute directly instead to the mass term $\frac{1}{2} M_{\Sigma}\, \chi \chi$; note that this  mass corresponds to the operator
\beq
\C{L} \supset \int d^4 \theta \frac{{\mathbf X}^\dagger}{2M} {\mathbf \Sigma}^2 + h.c. \rightarrow \int d^2 \theta \frac{F^\dagger}{M} \frac{1}{2}{\mathbf \Sigma}^2 + h.c.
\eeq
i.e. it is generated via the Giudice-Masiero mechanism \cite{Giudice:1988yz} and can be written as a supersymmetric mass term.
\beq
M_{\Sigma} = \frac{1}{16\pi^2} \frac{F^{\dagger}}{\Mm} \mathrm{tr} (\kappa^\dagger \lambda^2)
\eeq
at leading order. Thus, we require:

\beq
\mathrm{tr} (\kappa^\dagger \lambda^2) =0.
\eeq

\subsection{No large D term contribution to Soft Masses}

Kinetic mixing 
leads to a hypercharge D term and thus a contribution to the soft masses. Integrating out messengers at a scale $\Lambda$ we find the holomorphic kinetic mixing $\chi_h$ given by
\beq
\chi_h (\Lambda) = - \frac{1}{8\pi^2} \tr \bigg( Q Q' \log \C{M}/\Lambda\bigg)
\eeq
assuming there are no massless states charged under both gauge groups. 
The term in the Lagrangian is
\beq
\C{L} \supset \int d^2 \theta -\frac{1}{2} \chi W^\alpha W^\prime_\alpha +c.c  \supset  -\frac{1}{2} D D^\prime (\chi + \ov{\chi}).\eeq 
This is to be compared to $\int d^4 \theta 2 \xi V \supset - \xi D$, giving $\xi =  \frac{1}{2} D^\prime (\chi + \ov{\chi})$. Since the mixing that should appear is the physical one \cite{Benakli:2009mk}, we have corrections to the sfermion masses $ \Delta m_{\tilde{f}}^2$ given by
\begin{align}
\Delta m_{\tilde{f}}^2 &= g_Y^2 Y_f \xi \nonumber \\
&=-g_Y^3 Y_f  g^\prime D^\prime \bigg[ \Re (\chi_h) + \frac{1}{8\pi^2} \mathrm{tr}\bigg( Q Q^\prime \log Z \bigg) \bigg] \nonumber \\
&=-g_Y^3 Y_f  g^\prime D^\prime \frac{1}{16\pi^2} \tr \bigg( Q Q' \log |\C{M}|^2/\Lambda^2\bigg).
\end{align}
This yields 
\beq
\Delta m_{\tilde{f}}^2 = -g_Y^3 Y_f  g^\prime D^\prime \frac{1}{8\pi^2} \sum_r  2 \tr ( \hat{e} \hat{Y}) \log M^r/\Lambda .
\eeq

This is a potentially extremely dangerous term, and we must therefore ensure that 
\beq
\tr ( \hat{e} \hat{Y}) = 0 
\eeq
for each set of messengers separately. 

\subsection{No large tadpole terms for the adjoint scalars}

In addition the absence of tachyons for the adjoint scalars, we also require the absence of linear couplings $(D^2, F^2)\times (\Sigma+\Sigma^\dagger)$. The term in the K\"ahler potential for the $F$-term spurion is
\beq
K \supset -\frac{|\mathbf{X}|^2}{32\pi^2 \Mm} \bigg[ \Sigma \tr (\lambda \{\kappa, \kappa^\dagger\}) + \Sigma^\dagger \tr (\lambda \{\kappa, \kappa^\dagger\})\bigg]
\eeq
and thus the dangerous terms in the potential are
\begin{align}
V \supset& \frac{|F|^2}{32\pi^2 \Mm} \bigg[ \Sigma \tr (\lambda \{\kappa, \kappa^\dagger\}) + \Sigma^\dagger \tr (\lambda^\dagger \{\kappa, \kappa^\dagger\})\bigg] \nonumber \\
&+\frac{D^2}{16\pi^2 \Mm} \tr (\Sigma \lambda \hat{e}^2 + \Sigma^\dagger \lambda^\dagger \hat{e}^2).
\end{align}
Since the mass terms are $\C{O}(\lambda^2 F^2/\Mm)$, these linear terms would lead to a vev for the $U(1)$ adjoint of $\C{O}(\Mm/\lambda)$, and thus since the $U(1)$ adjoint couples to all of the messengers the above must always be zero. We cannot hope to have cancellations between the terms, since this would require extreme fine tuning of the parameters; we must impose
\begin{align}
\tr (\lambda \{\kappa, \kappa^\dagger\}) &=0 \nonumber \\
\tr (\lambda \hat{e}^2) &=0.
\end{align}
The latter condition implies that 
\beq
\tr(\lambda) = 0  \qquad \mathrm{or} \qquad \hat{e}^2 = 0.
\eeq

\subsection{Summary}

In summary, the following constraints must be applied to all matrices:
\begin{align}
\mathrm{tr} (\kappa)=& 0 \nonumber \\
\mathrm{tr} (\kappa^\dagger \lambda^2) =&0\nonumber \\
\tr ( \hat{e} \hat{Y}) =& 0\nonumber \\
\tr (\lambda \{\kappa, \kappa^\dagger\}) =&0\nonumber \\
\tr (\lambda \hat{e}^2) =&0.
\end{align}
Moreover, to give leading order Dirac gaugino masses via D-terms, we require 
\beq
\mathrm{tr} (\lambda \hat{e}) \ne 0.
\eeq
This must always apply to the $SU(2)$ and $SU(3)$ couplings, but since the singlet adjoint may couple to any set of messengers this constraint is less restrictive in that case. 

We must also avoid tachyonic scalars (at least for the $SU(2)$ and $SU(3)$ adjoints) and thus we require $ m_\Sigma^2 \ge B_\Sigma $, which constrains the coupling matrices via equations (\ref{AdjointDtermMasses}) and (\ref{AdjointFtermMasses}).

\section{Renormalisation}
\label{RENORMALISATION}

Running the parameters of a theory from a high energy scale down to low energies can have a profound effect upon their values; it is thus important to consider how all the parameters in the theory run. In this work we present the renormalisation group equations for our model to one loop order, postponing the extension to two loops for future work. In this section we consider first the messenger couplings, then  the superpotential couplings, and finally the soft parameters. We give here the RGEs in their simplest practical form, with third-generation dominant running and assuming an R-symmetry that is only broken by an explicit $B_\mu$ term; the subsequent running does not generate $A$ terms or Majorana masses at one loop order. The general RGEs allowing Majorana masses and three-generation running are presented in appendix \ref{Appendix:RGEs}.

\subsection{Running Messenger Couplings}

Recalling the couplings between the adjoints and the messengers
\beq
W \supset \lambda^{(ab)}_{i\ov{j}} \mathrm{tr}(\mathbf{Q_{ia} \Sigma_b \tilde{Q}_{\ov{j}a}}) 
\eeq
we must take into account that these run; the RGE for this coupling is
\begin{align}
16\pi^2 \frac{d \lambda_{i \tilde{j}}^{(ab)}}{dt} =&  \lambda^{(ab)}_{i \tilde{j}} \bigg[ \sum_c I^b_{Q_{ic}} \tr (\lambda^{(cb)} (\lambda^{(cb)})^\dagger ) -2g^2_b [ 2 C_{Q_{ia}}^b + C_{G^b}^b ]  \nonumber \\
&\qquad\qquad + \frac{1}{2} \sum_{mn} Y^{\Sigma_b mn} Y_{\Sigma_b mn}\bigg] \nonumber \\
& +   \sum_c C_{Q_{ia}}^c  \bigg[ \lambda^{(ac)} (\lambda^{(ac)})^\dagger \lambda^{(ab)} + \lambda^{(ab)} \lambda^{(ac)} (\lambda^{(ac)})^\dagger \bigg] ,
\end{align}
where $C_{G^b}^b$ is the Casimir of the group $G^b$ (so $N$ for $SU(N)$), and the term \\ $ \frac{1}{2}\sum_{mn} Y^{\Sigma_b mn} Y_{\Sigma_b mn}$ is summing over all additional couplings of the adjoints (we are necessarily neglecting any hidden-sector couplings of the messengers). Clearly the terms $\lambda \lambda^\dagger \lambda$ can mix the components; other than the obvious aesthetic defect that this causes, the change of the relative size of the entries in this matrix would be very dangerous in allowing terms such as adjoint singlet tadpoles or hypercharge $D$-terms to be generated, even if we had chosen the couplings to avoid this at one loop - we can see this since  the renormalisation of the messenger mass matrix is 
\beq
\frac{d}{dt} M^{(a) i \tilde{j}}_{\mathrm{Mess}} = \sum_b \frac{C_{Q_{ia}}^b}{16\pi^2}  [ \lambda^{(ab)} (\lambda^{(ab)})^\dagger M^{(a)}_{\mathrm{Mess}} + M^{(a)}_{\mathrm{Mess}} \lambda^{(ab)} (\lambda^{(ab)})^\dagger - 4 g^2_b M^{(a)}_{\mathrm{Mess}} ]^{i \tilde{j}}.
\eeq  
However, if we choose $ \lambda^{(ab)}_{i\ov{j}}$ to be proportional to a unitary matrix, or zero, then mass splittings can be avoided;  suppose we split the couplings into block-diagonal segments   $ \lambda^{(ab)}_{i\ov{j}} = y_{(ab)}^k u_{(ab)i\ov{j}}^k$, with $\tr(u_{(ab)}^{k} (u_{(ab)}^l)^\dagger) \equiv \delta^{kl} N_{(ab)}^k, u_{(ab)}^{k} (u_{(ab)}^l)^\dagger \equiv \delta^{kl} \eta_{(ab)k} $, with $\eta_{(ab)k}$ equal to zero or one. Then the RGEs become
\begin{align}
16\pi^2 \frac{d \log y_{(ab)}^k}{dt} =&    -2g^2_b [ 2 C_{Q_{ia}}^b + C_{G^b}^b ]   + \frac{1}{2} \sum_{mn} \tr (Y^{\Sigma_b mn} Y_{\Sigma_b mn}) \nonumber \\
& +  \sum_c  2 C_{Q_{ia}}^c \eta_{(ac)k} (y_{(ac)}^k)^2 + \sum_d\sum_l I_{Q_{id}}^b  N_{(db)}^l (y_{(db)}^l)^2 .
\end{align}

\subsection{Supersymmetric Couplings}

As the Higgs doublets have new superpotential couplings to the DG-adjoints, the Yukawa couplings $y_t, y_b, y_\tau$ of the  top, bottom and tau fermions become:
\begin{eqnarray}
\frac{d}{dt} y_t &=&  \frac{y_t}{16\pi^2} \bigg[ \lambda_S^2 + 3\lambda_T^2 + 6 |y_t|^2 +  |y_b|^2 - \frac{16}{3} g_3^2  - 3g_2^2 - \frac{13}{9} g_Y^2 \bigg] \nonumber \\
\frac{d}{dt} y_b &=&  \frac{y_b}{16\pi^2} \bigg[ \lambda_S^2 + 3\lambda_T^2 + 6 |y_b|^2 +  |y_t|^2 + |y_{\tau}|^2 - \frac{16}{3} g_3^2  - 3g_2^2 - \frac{7}{9} g_Y^2 \bigg] \nonumber \\
\frac{d}{dt} y_\tau &=&  \frac{y_\tau}{16\pi^2} \bigg[ \lambda_S^2 + 3\lambda_T^2 + 4 |y_\tau|^2  + 3|y_b|^2  - 3g_2^2 - 3 g_Y^2 \bigg] \end{eqnarray}

The running of the couplings of the Higgs to the adjoints is given by
\begin{align}
16\pi^2 \frac{d \log \lambda_S}{dt} =& -g_Y^2 -3g_2^2 +4 \lambda_S^2 + 6\lambda_T^2 + 3 |y_t|^2 + 3 |y_b|^2 + |y_{\tau}|^2 \nonumber \\
& + \sum_c \sum_l Y_{c}^2 N_{(cY)}^l (y_{(cY)}^l)^2 \nonumber \\
16\pi^2 \frac{d \log \lambda_T}{dt} =&-g_Y^2 -7g_2^2 +2 \lambda_S^2 + 8\lambda_T^2 + 3 |y_t|^2 + 3 |y_b|^2 + |y_{\tau}|^2 \nonumber \\
&+  \sum_c \sum_l I^2_{Q_{ic}} N_{(c2)}^l (y_{(c2)}^l)^2.
\end{align}

\subsection{Soft Parameters with Dirac Gaugino Masses}

As discussed in section \ref{LOWENERGY}, the presence of adjoint superfields allows the generation of non-standard soft supersymmetry breaking terms, which modify the renormalisation group equations. General nonstandard soft terms greatly complicate the renormalisation group equations, as independent $r$ parameters introduce a very large number of new equations; the RGEs for generic models allowing such terms are presented in \cite{Jack:1999ud,Jack:1999fa}. However, in our model, since supersymmetry is broken spontaneously and only has nonstandard terms arising from a superpotential coupling the relation (\ref{req}) is obeyed even under renormalisation group running; we thus do not give independent renormalisation group equations for these parameters. Moreover, the supersoft nature of such operators allows their impact upon the RGEs for the standard soft terms to be easily determined through substitution $m_{\Sigma^b}^2 \rightarrow m_{\Sigma^b}^2 - 2 m_{bD}^2$ and $B_{\Sigma^b} \rightarrow B_{\Sigma^b} - 2m_{bD}^2$ in the equations for standard soft supersymmetry breaking terms \cite{Jack:1999fa}.

The equation for the tadpole term is given by
\beq
\frac{d}{dt} t^S = \frac{1}{16\pi^2} \bigg[ 2 \lambda_S^2 t^S + 4 \lambda_S \mu (m^2_{H_d} + m^2_{H_u}) + 2\sqrt{2} g_Y m_{1D} \mathrm{Tr}(Y m^2) \bigg].
\eeq

The Dirac gaugino masses run as:
\begin{eqnarray}
\frac{d}{dt} m_{1D} &=&  \frac{m_{1D}}{16\pi^2} \bigg[ 11 g_Y^2 + 2\lambda_S^2 \bigg] \nonumber \\ 
\frac{d}{dt} m_{2D} &=&  \frac{m_{2D}}{16\pi^2} \bigg[ - g_2^2 + 2\lambda_T^2 \bigg] \nonumber \\ 
\frac{d}{dt} m_{3D} &=&  - \frac{m_{3D}}{16\pi^2} \times 6 g_3^2 .
\end{eqnarray}
The adjoint scalar mass equations are:
\begin{align}
\frac{d}{dt} m_S^2 =& \frac{1}{16\pi^2} \bigg[ 4\lambda_S^2 [m^2_{H_u} + m^2_{H_d} + m_{S}^2] + 44 g_Y^2 m_{1D}^2\bigg] \nonumber \\
\frac{d}{dt} m_T^2 =& \frac{1}{16\pi^2} \bigg[ 4\lambda_T^2 [m^2_{H_u} + m^2_{H_d} +  m_{T}^2] - 4 g_2^2 m_{2D}^2 \bigg] \nonumber \\
\frac{d}{dt} m_O^2 =& \frac{1}{16\pi^2} \bigg[ -24 g_3^2 m_{3D}^2 \bigg] \nonumber \\
\frac{d}{dt} B_S =& \frac{1}{16\pi^2} \bigg[ 4 \lambda_S^2 B_S + 44 g_Y^2 m_{1D}^2 \bigg] \nonumber \\
\frac{d}{dt} B_T =& \frac{1}{16\pi^2} \bigg[ 4 \lambda_T^2 B_T - 8 g_2^2 B_T + 12 g_2^2 m_{2D}^2 \bigg]  \nonumber \\ 
\frac{d}{dt} B_O =& \frac{1}{16\pi^2} \bigg[  - 12 g_3^2 B_O  \bigg]
\label{AdjointMassRunning}\end{align}

The soft terms for the Higgs run as
\begin{align}
16\pi^2 \frac{d}{dt} m^2_{H_u} =&  6|y_t|^2 [m^2_{Q_3} + m^2_{U_3} + m^2_{H_u}] \nonumber \\
& + 2 \lambda_S^2 [m^2_{H_u} + m^2_S + m^2_{H_d}]  + 6 \lambda_T^2 [m^2_{H_u} + m^2_T + m^2_{H_d}] \nonumber \\
& +  g_Y^2  Tr(Y m^2) \nonumber \\ 
&- 4 \lambda_S^2 m_{D1}^2 - 12 \lambda_T^2 m_{D2}^2 \nonumber \\
16\pi^2 \frac{d}{dt} m^2_{H_d} =&  6|y_b|^2 [m^2_{Q_3} + m^2_{D_3} + m^2_{H_d}] \nonumber \\
& + 2|y_\tau|^2 [m^2_{L_3} + m^2_{E_3} + m^2_{H_d}] \nonumber\\
& + 2 \lambda_S^2 [m^2_{H_u} + m^2_S + m^2_{H_d}]  + 6 \lambda_T^2 [m^2_{H_u} + m^2_T + m^2_{H_d}] \nonumber \\
& -  g_Y^2  Tr(Y m^2)  \nonumber \\
&- 4 \lambda_S^2 m_{D1}^2 - 12 \lambda_T^2 m_{D2}^2  \nonumber \\
16\pi^2 \frac{d}{dt} B_\mu =&  B_\mu[3 |y_t|^2 +  3 |y_b|^2 +|y_\tau|^2 - 3 g_2^2 - y_Y^2] \nonumber \\
& + 2 B_\mu \lambda_S^2 + 6 B_\mu \lambda_T^2 
\end{align}

Defining as usual
\begin{align}
X_t \equiv& 2|y_t|^2(m_{H_u}^2 + m_{Q_3}^2 + m_{U_3}^2)  \nonumber \\
X_b \equiv& 2|y_b|^2(m_{H_d}^2 + m_{Q_3}^2 + m_{D_3}^2) \nonumber \\
X_\tau \equiv& 2|y_\tau|^2(m_{H_d}^2 + m_{L_3}^2 + m_{E_3}^2)  
\end{align}
we have the sfermion mass equations
\begin{align}
16\pi^2 \frac{d}{dt} m^2_{Q_3} =& X_t + X_b + \frac{1}{3} g_Y^2  Tr(Y m^2)   \nonumber \\
16\pi^2\frac{d}{dt} m^2_{U_3} =&  2X_t  - \frac{4}{3} g_Y^2  Tr(Y m^2)   \nonumber \\
16\pi^2\frac{d}{dt} m^2_{D_3} =&  2X_b+ \frac{2}{3} g_Y^2  Tr(Y m^2)  \nonumber \\
16\pi^2\frac{d}{dt} m^2_{L_3} =&  X_\tau  - g_Y^2  Tr(Y m^2)   \nonumber \\
16\pi^2\frac{d}{dt} m^2_{E_3} =&  2X_\tau   + 2 g_Y^2  Tr(Y m^2) .
\end{align}

\section{Achieving unification of gauge couplings}
\label{UNIFICATION}

We require that the gauge couplings remain perturbative up to a high scale, $M_{U}$, the unification scale, to keep calculability in the whole energy range. Unification is assumed to appear as a relation existing between the gauge couplings at  $M_U$ reflecting the way the gauge interactions are unified in the considered framework. We shall consider   the peculiar equality predicted in Grand Unified Theories, with a normalization of the hypercharge $U(1)_Y$ corresponding to an $SU(5)$ embedding. Other possibilities exist such as equality up to integer multiplicative factors, the Kac-Moody levels, in perturbative heterotic strings, but these shall not be considered here.

The introduction of adjoint superfields at a low energy scale requires additional states introduced at an intermediate scale in order to preserve unification. We would then like to be able to identify these states as messengers of gauge mediation of supersymmetry breaking. This excludes many possible set of states such as the ``bachelor field'' examples discussed in \cite{Fox:2002bu}, including  embedding the adjoints in $\mathbf{24} = (\mathbf{8},1)_0 + (1,\mathbf{3})_0 + (\mathbf{1},\mathbf{1})_0 + (\mathbf{3},\mathbf{\ov{2}})_{-5/6} + (\mathbf{\ov{3}},\mathbf{2})_{5/6}$ of $SU(5)$; without introducing $R$-symmetry breaking supersymmetric adjoint masses these lead to tachyonic adjoint scalars.

As input, we take the data  at the electroweak scale:
\begin{eqnarray}
m_Z = 91.18 \  \  {\rm GeV},&& \qquad 
\alpha_{em}(m_Z)= \frac{1}{127.9},  \qquad \sin^2{\theta_w}(m_Z) = 0.231 \nonumber  \\
\frac{ \alpha_1 }{k_1} \equiv  \alpha_Y  =  \frac{\alpha_{em}}{\cos^2{\theta_w}} , && \qquad
\alpha_2 =  \frac{\alpha_{em} }{\sin^2{\theta_w}}, \qquad {\rm and} \quad  \alpha_{3}(m_Z)=0.1187
\end{eqnarray}
the running quantities being defined in the {\={MS}} scheme.  In the MSSM, at leading order, neglecting running between $m_Z$ and the supersymmetric soft masses average mass scale, the gauge couplings run as
\beq
\frac{1}{\alpha_i (\mu)} = \frac{1}{\alpha_i (M)} - \frac{b_i}{2\pi} \log \mu/M+ \Delta_i
\eeq
where the beta function coefficients are $b_i = (-3,1,11/k_1)$. Here we have left the normalization of the hypercharge $U(1)_Y$ arbitrary, labeled by $k_1$. For $k_1=5/3$ which corresponds to embedding in $SU(5)$, this is compatible with unification at a scale  $M_U \sim 3\cdot 10^{16}$GeV and a coupling $\alpha_U \sim 1/24$.

For the Dirac gaugino masses to be relevant, the masses of the DG-adjoint are to be quite light. Thus, they contribute to the running of the gauge coupling all the way up to the unification scale. In fact, the inclusion of the DG-adjoints modifies the beta function coefficients of $SU(2)$ and $SU(3)$ $b_{20}$ and $b_{30}$ respectively, and unless new states are also present unification is lost.
The set of new fields is taken to have the smallest contribution to the running, and thus fall into fundamental representations; to be vector-like to allow arbitrary masses; to come in more than one pair for each group to allow the use as messengers for generating Dirac gaugino masses without tachyonic adjoint scalars.  This set is parametrized as  $n_1$ pairs of multiplets $[(1,1)_{ y_1}+ (1,1)_{- y_1}]$ with mass $m_1$ , $n_2$ pairs of multiplets $[(1,2)_{y_2}+(1,2)_{-y_2}]$ with mass $m_2$ and $n_3$ pairs of multiplets $[(3,1)_{ y_3}+(3,1)_{ -y_3}]$ with mass $m_3$, where the numbers between parenthesis are the  $SU(3)$ and $SU(2)$  quantum numbers respectively, and $y_i$ are the associated hypercharges. In order to keep the couplings perturbative up to very high energies, the numbers $n_1, n_2, n_3$ are to be small, and the masses $m_i$  large.   We will consider for instance the case $m_Z< m_3, m_2, m_1 <\mu$, and the couplings at the scale $\mu$ are related to the ones at $m_Z$ by the relations:

\begin{align}
\frac{1}{  \alpha_1 (\mu)}=&\frac{1}{k_1 \alpha_Y (m_Z)} -\frac{b_{10}}{2\pi }\log  \left[\frac{\mu}{m_Z}\right]-\frac{b_{13}}{2\pi } \log  \left[\frac{\mu}{m_3}\right] -  {\frac{b_{12}}{2\pi } \log \left[\frac{\mu}{m_2}\right]-\frac{b_{11}}{2\pi } \log \left[\frac{\mu}{m_1}\right]} \nonumber \\
\frac{1}{\alpha_2 (\mu)} =& \frac{1}{\alpha_2 (m_Z)}  - \frac {b_{20}}{2\pi } \log  \left[\frac{m_2}{m_Z}\right]-\frac{b_{21}}{2\pi } \log \left[\frac{\mu}{m_2}\right] \nonumber \\
\frac{1}{\alpha_3 (\mu)} =&\frac{1}{\alpha_3 (m_Z)}  - \frac {b_{30}}{2\pi } \log  \left[\frac{m_3}{m_Z}\right]-\frac{b_{31}}{2\pi } \log \left[\frac{\mu}{m_3}\right] \end{align}

where the beta function coefficients are given by
\begin{eqnarray}
b_{10}=11/k_1\qquad  b_{20}=1 +2=3 \qquad && b_{30}=-3+3=0 \nonumber \\
b_{13}=(6 {n_3} {y_3}^2)/{k_1}  \qquad  
b_{12}= (4{n_2} {y_2}^2)/{k_1}\qquad  && 
b_{11}=(2{n_1} {y_1}^2)/ {k_1} \nonumber  \\
b_{21}=b_{20}+ {n_2}\qquad 
b_{31}=b_{30}+ {n_3}&&
\end{eqnarray}
As stated above, we choose   $k_1=5/3$, and require equality of the gauge couplings $\alpha_1 = \alpha_2=\alpha_3= \alpha_U $  at $M_U $. There many possible sets that lead to this unification, but we are interested in those with a minimal additional states, and for this paper a unification scale close to the heterotic string prefered value of order $\sim 2 \sqrt{\alpha_U}\cdot 10^{18}$ GeV, as that will mean also unification with the gravitational interactions. We shall  keep a minimal spread of the masses of the messengers.

The smallest value of $\alpha_U $ is obtained when no coloured state is added. Without doublets, the $SU(2)$ and $SU(3)$ couplings meet at energies above $10^{18} {\rm GeV}$.   Including the minimal number of triplets to generate the required masses for the $SU(3)$ DG-adjoints, one can find different solutions. For instance, we choose 
\begin{eqnarray}
4\times [(1,1)_{ 1}+ (1,1)_{- 1}]  &&\qquad  {\rm at}  \qquad   m_1=  3 \ 10^{12} {\rm GeV} \nonumber  \\
4\times [(1,2)_{ 1/2}+ (1,1)_{- 1/2}]  &&\qquad  {\rm at}  \qquad   m_2=   1.3\  10^{13} {\rm GeV} \nonumber  \\
2\times [(3,1)_{ 1/3}+ (3,1)_{- 1/3}]  &&\qquad  {\rm at}  \qquad   m_3= 10^{13} {\rm GeV}  \nonumber  \\
M_U\sim 9.9 \cdot  10^{17} {\rm GeV} \qquad  &&\qquad  \qquad \qquad   \alpha_U^{-1} \sim 4.77
\end{eqnarray}
where we have a set of unifying scale and coupling in agreement with the heterotic string leading order prediction. The small mass splitting allows perfect equality of couplings at the GUT scale; this splitting could be due to running effects or echanged for threshold corrections. 


\section{Model Building}
\label{MODELBUILDING}

\subsection{Avoiding tachyonic adjoint scalars, part II}

As can be seen from equation (\ref{AdjointMassRunning}), the strong coupling amplifies the $B_O$ term much more than the $m_O$ term, and so we find on running from a high messenger scale that unless $B_O$ is much smaller than $m_O^2$ (for the messenger masses that we shall consider for explicit models below, at least a factor of ten is required) the octet adjoints become tachyonic at low energies. We can however solve this problem by judicious choice of the adjoint-messenger coupling matrix $u_{O}$; a particular class of choices 
\begin{equation}
u = \C{V} (x,\theta) \equiv \frac{1}{\sqrt{4x^2 - 2}} \2b2[1+ix,e^{i\theta} \sqrt{3(x^2-1)}][e^{-i\theta} \sqrt{3(x^2-1)},-1+ix] .
\end{equation}
Here $x^2 > 1$ is a real number and $\theta$ is a phase, which may be chosen to be zero. Note that this generates a \emph{real} mass for both the gauginos and the adjoints; the adjoint mass is 
\begin{equation}
V \supset 2^{-\delta} \frac{D^2}{\Mm^2} \frac{ y^2 }{32\pi^2} 8 \left( \frac{x^2-1}{2x^2-1} \right) \Sigma^a (\Sigma^a)^* ,
\end{equation}
i.e. $B_\Sigma=0$ (recall $\delta = 1$ for $SU(N)$, $0$ for $U(1)$ adjoints). Since there is no mass generated for the operator $(\Sigma^a)^2$, we will not generate a tachyon upon running to low energies. 

Note that the requirement of absence of tadpoles for the singlet scalar means that, if we wish to generate Dirac masses for the Bino, the singlet chiral superfield must have different couplings to the messengers. A form of particular interest is
\beq
u = \C{U} (x) \equiv \frac{1}{\sqrt{1+x^2}} \2b2[1,ix][-ix,-1],
\eeq
where $x$ is again a real parameter. This has a unitary structure, will generate Dirac gaugino masses and is traceless - so will not generate tadpoles. The adjoint scalar masses generated by $D$-terms are \cite{Benakli:2008pg}:
\begin{align}
V \supset& \frac{|y|^2 D^2}{32\pi^2 \Mm^2} \frac{2}{3} \bigg( \frac{ 4 x^2 }{1+x^2} |S|^2 - \frac{3+x^2}{1+x^2} (S^2 + \ov{S}^2 ) \bigg) \nonumber \\
m_S^2 =& \frac{|y|^2 D^2}{16\pi^2 \Mm^2} \frac{1}{3} \frac{ 4 x^2 }{1+x^2} \nonumber \\
B_S =&  - \frac{|y|^2 D^2}{16\pi^2 \Mm^2} \frac{2}{3} \frac{3+x^2}{1+x^2}
\end{align}
and thus
\beq
m_{S_P,_M}^2 = \frac{|y|^2 D^2}{24\pi^2 \Mm^2} \frac{2 x^2 \mp (x^2 + 3)}{1+x^2}.
\eeq

\subsection{Explicit Models}

Here we finally assemble the ingredients in the previous sections to present some explicit models of gauge mediation with Dirac gauginos and unified gauge couplings. We shall take three sets of messenger superfield pairs $\mathbf{Q_{ia}}, \mathbf{ \tilde{Q}_{\ov{j}a}}$, as given in table \ref{diracgauginos_Fields}, in representations of $SU(3) \times SU(2) \times U(1)_Y$ as follows: $2 \times [(\mathbf{3},\mathbf{1},\mathbf{-1/3}), (\mathbf{\bar{3}},\mathbf{1},\mathbf{1/3})]$ at mass $M_{\mathrm{Mess}}^{(3)} = 10^{13}$ GeV; $4 \times [(\mathbf{1},\mathbf{2},\mathbf{1/2}), (\mathbf{1},\mathbf{\bar{2}},\mathbf{-1/2})]$ at mass $M_{\mathrm{Mess}}^{(2)} = 1.3 \times 10^{13}$ GeV and $4 \times [(\mathbf{1},\mathbf{1},\mathbf{1}), (\mathbf{1},\mathbf{1},\mathbf{-1})]$ at mass $M_{\mathrm{Mess}}^{(1)} = 3 \times 10^{12}$ GeV. These masses are defined to be their values at the messenger scale.

Our specific choice  matrices for the couplings of the adjoints to the messengers $W \supset \lambda^{(ab)}_{i \ov{j}} (\mathbf{Q_{ia} \Sigma^b \tilde{Q}_{\ov{j}a}})$ is
\begin{align}
\lambda^{(S1)} = & 0 \nonumber \\
\lambda^{(S2)} = & y_{S2} \,\mathrm{diag}( \C{U}(x_{U}),0) \nonumber \\
\lambda^{(S3)} =& - y_{S3} \,\C{U} (x_{U}) \nonumber \\
\lambda^{(T2)} =& \left(\begin{array}{cc} y_{T}^{(1)} \,\C{V} (x_V,0) & 0\\0 &y_T^{(2)}\sigma_3 \end{array}\right)  \nonumber \\
\lambda^{(O3)} =& y_O \C{V} (x_V,0) ,
\end{align}
i.e. the $\mathbf{Q_{i1}}$ messengers do not couple to the adjoints, and the $\mathbf{Q_{i2}}$ messengers are split into two blocks of two; we shall set $y_T^{(1)} = y_T^{(2)}$ at the GUT scale (although they do acquire small differences at the messenger scale).

The couplings to the $F$- and $D$-terms are given by
\begin{align}
\kappa^{(1)} =& \mathrm{diag}(\sigma^+,\sigma^+) \nonumber \\ 
\kappa^{(2)} =& \mathrm{diag}(0,\sigma^+) \nonumber \\
\kappa^{(3)} =&0 \nonumber \\
\hat{e}^{(1)} =& 0 \nonumber \\
\hat{e}^{(2)} =& \mathrm{diag} (\sigma_3,0) \nonumber \\
\hat{e}^{(3)} =& \sigma_3 
\end{align}
where $\sigma^+ = \2b2[0,1][0,0], \sigma_3 = \2b2[1,0][0,-1]$ are the usual Pauli matrices. This framework then allows the generation of a large variety of spectra; we illustrate for three cases in tables \ref{Model_Parameters} and \ref{softmasses}. In each case, there is some unification of the new couplings. $\lambda_S/\lambda_T = \sqrt{3/5} , y_{S2}/y_{S3} = 3/2 $  in all three models, although $\lambda_S, \lambda_T$ are smaller than the $N=2$ value ($\sqrt{2} g (M_{GUT})$) in each case. In models I and II further $y_2/y_3 = 1$ at the GUT scale, and furthermore in model II all the messenger couplings unify at the ``N=2'' values (of course, they do not fall into true $N=2$ hypermultiplets since the couplings are not diagonal). The first two models have a relatively small $B_\mu$ term, model II having also a small $\mu$-term. The chief virtue of model I is a light bino, but if the gravitino is the LSP (i.e. if SUSY-breaking gravity effects are not sufficiently sequestered) then the wino may be too light; the other two models evade this restriction. Model III has light octet scalars which could provide interesting signatures at the LHC \cite{DGatLHC}.

\begin{table}[!h]
\begin{center}
\begin{tabular}{|c||c|c||c|c||c|c|}\hline
&\multicolumn{2}{c||}{Model-I}& \multicolumn{2}{c||}{Model-II} & \multicolumn{2}{c|}{Model-III}   \\ \hline  \hline 
Parameter &\multicolumn{6}{c|}{Input}  \\ \hline
$F( \mathrm{GeV}^2)$ &\multicolumn{2}{c||}{ $7.5 \times 10^{17}$}    &\multicolumn{2}{c||}{ $5.5 \times 10^{17}$}   &\multicolumn{2}{c|}{ $1.3 \times 10^{18}$}    \\ 
$D ( \mathrm{GeV}^2)$ &\multicolumn{2}{c||}{ $7.5 \times 10^{17}$}    &\multicolumn{2}{c||}{ $5.5 \times 10^{17}$}   &\multicolumn{2}{c|}{ $1.1 \times 10^{18}$}    \\ 
$x_{U}$ &\multicolumn{2}{c||}{ 2}    &\multicolumn{2}{c||}{ 1.9}   &\multicolumn{2}{c|}{ 2}    \\ 
$x_{V} $&\multicolumn{2}{c||}{ 1.5}     &\multicolumn{2}{c||}{ 1.1}& \multicolumn{2}{c|}{ 1}   \\
$y_{S1}$ &\multicolumn{2}{c||}{ 0}    &\multicolumn{2}{c||}{ 0}   &\multicolumn{2}{c|}{ 0}    \\ 
$y_{S2} $&\multicolumn{2}{c||}{ 0.317}     &\multicolumn{2}{c||}{ 0.709}& \multicolumn{2}{c|}{ 0.224}   \\ 
$y_{S3}$ & \multicolumn{2}{c||}{0.211}&\multicolumn{2}{c||}{ 0.473} &\multicolumn{2}{c|}{0.149}   \\ 
$y_{T}$ &  \multicolumn{2}{c||}{0.819} &  \multicolumn{2}{c||}{1.83}  & \multicolumn{2}{c|}{0.549}  \\ 
$y_{O}$ &  \multicolumn{2}{c||}{0.819}&   \multicolumn{2}{c||}{1.83}&  \multicolumn{2}{c|}{0.142}   \\  \hline  \hline 
& input & output& input & output& input & output   \\ \hline
$y_t$ & 0.32 &  0.993& 0.315 & 0.991 & 0.33 &0.991\\ 
$y_b$ & 0.16 &  0.691 & 0.158 & 0.688 &0.165 & 0.693 \\ 
$y_\tau$ & 0.2 & 0.295& 0.193 & 0.288 & 0.206 & 0.297\\ 
$\lambda_S$ & 0.0868& 0.0767& 0.0993 & 0.0769& 0.123 &0.106 \\ 
$\lambda_T$ & 0.112 & 0.152&0.128 &0.113 & 0.129 & 0.223 \\ 
$\mu (\mathrm{GeV})$&310&296& 101 & 98 & 330 & 301 \, \\
$B_\mu ( \mathrm{GeV}^2)$& -4490& -4320& -2209& -2180& -18200 & -16400\,\\ \hline 
\hline 
&\multicolumn{6}{c|}{Output}  \\ \hline
$\tan \beta$ & \multicolumn{2}{c||}{28.7} & \multicolumn{2}{c||}{ 28.6} &   \multicolumn{2}{c|}{ 28.8} \\ 
$\Delta \rho$&  \multicolumn{2}{c||}{ $ 2.18 \times 10^{-6}$}&   \multicolumn{2}{c||}{ $7.67 \times 10^{-5}$}&  \multicolumn{2}{c|}{ 0.000525}\\  \hline 
$\alpha_Y$& \multicolumn{6}{c|}{ 0.0105}\\ 
$\alpha_2 $& \multicolumn{6}{c|}{0.0332} \\ 
$\alpha_3 $& \multicolumn{6}{c|}{0.092}  \\  \hline 
\end{tabular}
\caption{Model parameters. }
\label{Model_Parameters}
\end{center}
\end{table}

\begin{table}[!h]
\begin{center}
\begin{tabular}{|c|c|c|c|c|}\hline
$\mathrm{Field}$&$\mathrm{Model-I}$&$\mathrm{ Model-II }$&$ \mathrm{Model-III}$\\ \hline  \hline
$m_{D1}$ & 127&134 &161   \\
$m_{D2}$ & 217& 308&472   \\
$m_{D3}$ & 1190&1710 &828  \\  \hline
$S_P$ & 1350&1100 & 1720  \\
$S_M $& 5320&5370 &6770   \\
$T_P $& 3590 &2190 & 1190  \\
$T_M $& 5890& 4910 &6500   \\
$O_P $& 5870&4020 & 1090  \\
$O_M $& 5870&4020 &1090   \\ \hline
$Q_3$ & 523&508 &442   \\ 
$Q_{1,2}$ & 617 &554 &791   \\ 
$U_3$ & 656 &583  &810   \\ 
$U_{1,2}$ & 786 &657 &1160   \\ 
$D_3$ & 477 &469  &369  \\ 
$D_{1,2} $& 535 & 504&587   \\ 
$L_3$ & 623 &459 & 1070   \\ 
$L_{1,2}$ & 652 & 480&1130     \\ 
$E_3$ & 956 &703 &1650   \\ 
$E_{1,2} $& 995 & 730& 1720  \\   \hline
$H_u$ & 308 i &127 i & 311 i  \\ 
$H_d $& 198 &237 &621   \\ \hline \hline
$A $& 352  &250 &689   \\ 
$h $& 117  &115 & 117  \\ 
$H$ & 351  & 248 & 692  \\  \hline
\end{tabular}
\caption{Low energy soft masses in GeV, with the exception that $A, h$ and $H$ are the physical Pseudoscalar, lightest scalar and heavy scalar Higgs masses respectively. }
\label{softmasses}
\end{center}
\end{table}

\section{Conclusions}
\label{CONCLUSIONS}

On one hand, achieving unification of gauge couplings is a very easy task, once new representations at intermediate scales are allowed. On the other hand, it is also quite easy to use  gauge mediation to construct models with (pseudo)-Dirac gauginos using the results of  \cite{Benakli:2008pg}. There, it was shown  how one can select sets of messengers and the associated superpotential that lead at the messenger scale to consistent models with Dirac gauginos.  However, the combination of  the two features  is not obvious. A tension appears due to the fact that, in order to generate sizable Dirac gaugino masses, one needs many messengers with low masses which tend to drive the gauge couplings quickly to become non-perturbative. Nevertheless, we have discussed how  such models can be constructed and have exhibited a few examples with explicit spectra at the electroweak scale.

In meeting this challenge, we have derived the set of constraints that need to be satisfied.  We have made use of simultaneous  contributions of  R-symmetric $F$ and $D$ terms. We have also found that the solution to the problem of tachyonic scalar adjoints of  
 \cite{Benakli:2008pg} (see also the example in \cite{Amigo:2008rc}) is not sufficient. The positive squared mass generated at  the messenger scale can be driven again to negative values at low energies by the renormalisation effects, in particular for the strongly coupled sgluons. We have solved this by a particular class of choices of messenger couplings that forbids the appearance of the superpotential contribution for their masses at one-loop, and as a result we have obtained degenerate masses for the real and imaginary components. While we have derived a few examples, a scan of the space of parameters needs to be performed in order to study the main features of the models. For example, we expect that a lower messenger scale would ameliorate this problem somewhat and allow a larger range of messenger couplings. 
 
To go further, an important problem  remains to be addressed. All our derivations of soft masses have an $R$-symmetric origin which protects against generation of Majorana gaugino masses. The model would have been $R$-symmetric if  not for the Higgs sector. A successful electroweak symmetry breaking, with a Higgs mass above the LEP limit, dictates the scale of $R$-symmetry breaking; in deriving our explicit models, we have taken $\mu$ and $B_\mu$ as parameters with values that satisfy the Higgs constraints. This procedure is common in models studying gauge mediation. However, in our case, it is even more important to understand the possible origin of such terms and to show that they do not induce a large contribution to Majorana gaugino masses. This question is under investigation.

\section*{Acknowledgments}

This work is supported in part by the European contract ``UNILHC'' PITN-GA-2009-237920. MDG is supported by the German Science Foundation (DFG) under SFB 676. MDG would like to thank the Laboratoire de Physique Th\'eorique et Hautes Energies (LPTHE) Paris, where part of this work was completed, for hospitality; and Andreas Ringwald and Thomas Underwood for stimulating discussions.

\appendix

\section{RGEs}
\label{Appendix:RGEs}

We present here the full three-family RGEs below the messenger scale, allowing both Dirac and Majorana gaugino masses, with MSSM $B_\mu$ and $A$ terms, supersymmetric adjoint masses, and $A_S, A_T, \kappa_S, A_{\kappa_S}$ nonzero.

The equation for the tadpole term is given by
\begin{align}
16\pi^2 \frac{d}{dt} t^S =&  (2 \lambda_S^2 + 2 |\kappa_S|^2) t^S + 2\sqrt{2} g_Y m_{1D} \mathrm{Tr}(Y m^2) + 4 \lambda_S \mu (m^2_{H_d} + m^2_{H_u})    \\
& + 4M_S \lambda_S B_\mu + 2 \kappa_S^* M_S B_S + 4 \kappa_S m_S^2 M_S^* + 4\lambda_S A_S B_\mu + 2 \kappa_SA_{\kappa_S} B_S^* . \nonumber
\end{align}

The Yukawa couplings of the  top, bottom and tau fermions (conventions given in equation \ref{Yukawa_Superpotential}) become:
\begin{align}
16\pi^2 \frac{d}{dt} Y_U =&   \bigg[  \lambda_S^2 + 3\lambda_T^2   + 3Y_U Y_U^\dagger  + 3 \mathrm{tr}(Y_U^\dagger Y_U) +  Y_D Y_D^\dagger - \frac{16}{3} g_3^2  - 3g_2^2 - \frac{13}{9} g_Y^2 \bigg] Y_U\nonumber \\
16\pi^2\frac{d}{dt} Y_D =&  \bigg[ \lambda_S^2 + 3\lambda_T^2  + 3Y_D Y_D^\dagger + 3 \mathrm{tr}(Y_D^\dagger Y_D)  \nonumber \\
&\qquad \qquad+  Y_U Y_U^\dagger + \mathrm{tr}(Y_E Y_E^\dagger)- \frac{16}{3} g_3^2  - 3g_2^2 - \frac{7}{9} g_Y^2 \bigg] Y_D \nonumber \\
16\pi^2\frac{d}{dt} Y_E =&  \bigg[ \lambda_S^2 + 3\lambda_T^2 + 3 Y_E Y_E^\dagger  +  \mathrm{tr}(Y_E Y_E^\dagger)  + 3 \mathrm{tr}(Y_D Y_D^\dagger)  - 3g_2^2 - 3 g_Y^2 \bigg] Y_E 
\end{align}
The running of the new couplings $\lambda_S, \lambda_T$ are 
\begin{align}
16\pi^2 \frac{d}{dt} \lambda_S =& \lambda_S \bigg[ 4 \lambda_S^2 + 6 \lambda_T^2 + 2 |\kappa_S|^2 + 3\tr(Y_U Y_U^\dagger) + 3\tr(Y_D Y_D^\dagger) + \tr(Y_E Y_E^\dagger) -  g_Y^2 - 3 g_2^2 \bigg] \nonumber \\
16\pi^2 \frac{d}{dt} \lambda_T =& \lambda_T \bigg[ 2 \lambda_S^2 +8 \lambda_T^2 + 3\tr(Y_U Y_U^\dagger) + 3\tr(Y_D Y_D^\dagger) + \tr(Y_E Y_E^\dagger) -  g_Y^2- 7 g_2^2\bigg] \nonumber \\
16\pi^2 \frac{d}{dt} \kappa_S =&   \kappa_S \bigg[6\lambda_S^2 + 6|\kappa_S|^2 \bigg].
\end{align}
The Dirac gaugino masses run as:
\begin{eqnarray}
\frac{d}{dt} m_{1D} &=&  \frac{m_{1D}}{16\pi^2} \bigg[ 11 g_Y^2 + 2\lambda_S^2 + 2|\kappa_S|^2\bigg] \nonumber \\ 
\frac{d}{dt} m_{2D} &=&  \frac{m_{2D}}{16\pi^2} \bigg[ - g_2^2 + 2\lambda_T^2 \bigg] \nonumber \\ 
\frac{d}{dt} m_{3D} &=&  - \frac{m_{3D}}{16\pi^2} \times 6 g_3^2 .
\end{eqnarray}
The adjoint fermion masses run as
\begin{align}
\frac{d}{dt} M_{S} =& \frac{M_{S}}{16\pi^2} \bigg[4 \lambda_S^2 +4 \kappa_S^2 \bigg]  \nonumber \\
\frac{d}{dt} M_{T} =& \frac{M_{T}}{16\pi^2} \bigg[ 4 \lambda_T^2 - 8 g_2^2 \bigg] \nonumber \\
\frac{d}{dt} M_{O} =&\frac{M_{O}}{16\pi^2} \times (-6) g_3^2 .
\end{align}
The adjoint scalar mass equations are:
\begin{align}
16\pi^2 \frac{d}{dt} m_S^2 =&  4\lambda_S^2 [m^2_{H_u} + m^2_{H_d} + m_{S}^2] + 44 g_Y^2 m_{1D}^2 + 16 \kappa_S^2 m_S^2 - 16 \kappa_S^2 m_{1D}^2  \nonumber \\
&+ 4 |A_S|^2 \lambda_S^2 + 4 \kappa_S^2 |A_{\kappa_S}|^2 \nonumber \\
16\pi^2\frac{d}{dt} m_T^2 =&  4\lambda_T^2 [m^2_{H_u} + m^2_{H_d} +  m_{T}^2] - 4g_2^2 m_{2D}^2  - 16 g_2^2 |M_2|^2 + 4 |A_T|^2 \lambda_T^2 \nonumber \\
16\pi^2\frac{d}{dt} m_O^2 =&  -24 g_3^2 m_{3D}^2  - 24 g_3^2 |M_3|^2 \nonumber \\
16\pi^2 \frac{d}{dt} B_S =&  (4 \lambda_S^2 + 4 |\kappa_S|^2) B_S + 44 g_Y^2 m_{1D}^2 + 8 \lambda_S^2 A_S M_S + 8 |\kappa_S|^2 A_{\kappa_S} M_S\nonumber \\
16\pi^2\frac{d}{dt} B_T =&  (4 \lambda_T^2  - 8 g_2^2 ) B_T + 12 g_2^2 m_{2D}^2 + 8 \lambda_T^2 A_T M_T  + 16 g_2^2 M_{T} M_2  \nonumber \\ 
16\pi^2\frac{d}{dt} B_O =& - 12 g_3^2 B_O + 24 g_3^2 M_{O} M_3 .
\end{align}

The Higgs masses run as
\begin{align}
16\pi^2 \frac{d}{dt} m^2_{H_u} =&  6\mathrm{tr} (Y_U Y_U^\dagger)m^2_{H_u} + 6 \mathrm{tr} ( Y_U^\dagger m^2_{Q} Y_U + Y_U m_U^2 Y_U^\dagger) \nonumber \\
& + 2 \lambda_S^2 [m^2_{H_u} + m^2_S + m^2_{H_d}]  + 6 \lambda_T^2 [m^2_{H_u} + m^2_T + m^2_{H_d}] \nonumber \\
& +  g_Y^2  Tr(Y m^2) \nonumber \\ 
&- 4 \lambda_S^2 m_{D1}^2 - 12 \lambda_T^2 m_{D2}^2 \nonumber \\
& + 6 \tr(A_U A_U^\dagger) + 2 |A_S|^2 \lambda_S^2 + 6 |A_T|^2 \lambda_T^2 - 2 g_Y^2 |M_1|^2 - 6 g_2^2 |M_2|^2 \nonumber \\
16\pi^2 \frac{d}{dt} m^2_{H_d} =& 6\mathrm{tr} (Y_D Y_D^\dagger)m^2_{H_d} + 6 \mathrm{tr} ( Y_D^\dagger m^2_{Q} Y_D + Y_D m^2_{D} Y_D^\dagger) \nonumber \\ 
& + 2\mathrm{tr} (Y_E Y_E^\dagger)m^2_{H_d} + 2 \mathrm{tr} ( Y_E^\dagger m^2_{L} Y_E + Y_E m^2_{E} Y_E^\dagger) \nonumber \\
& + 2 \lambda_S^2 [m^2_{H_u} + m^2_S + m^2_{H_d}]  + 6 \lambda_T^2 [m^2_{H_u} + m^2_T + m^2_{H_d}] \nonumber \\
& -  g_Y^2  Tr(Y m^2)  \nonumber \\
&- 4 \lambda_S^2 m_{D1}^2 - 12 \lambda_T^2 m_{D2}^2  \nonumber \\
& + 6 \tr(A_D A_D^\dagger) + 2 \tr(A_E A_E^\dagger) + 2 |A_S|^2 \lambda_S^2 + 6 |A_T|^2 \lambda_T^2 \nonumber \\
&- 2 g_Y^2 |M_1|^2 - 6 g_2^2 |M_2|^2 \nonumber \\
16\pi^2 \frac{d}{dt} B_\mu =&  B_\mu [3 \tr (Y_U Y_U^\dagger)  +  3 \tr (Y_D Y_D^\dagger) +\tr (Y_E Y_E^\dagger) - 3 g_2^2 - y_Y^2] \nonumber \\
& + \mu \,\tr( 6 A_U Y_U^\dagger + 6 A_D Y_D^\dagger +2 A_E Y_E^\dagger) + 4 \lambda_S^2 A_S \mu + 12 \lambda_T^2 A_T \mu\nonumber \\
& + 2 B_\mu \lambda_S^2 + 6 B_\mu \lambda_T^2 + 2 g_Y^2  M_1 \mu + 6 g_2^2 M_2 \mu 
\end{align}

The MSSM sfermion equations are identical to the MSSM ones:
\begin{align}
16\pi^2 \frac{d}{dt} m^2_{Q} =& (Y_U Y_U^\dagger m_Q^2) + (m_Q^2 Y_U Y_U^\dagger) + 2 (Y_U m_U^2 Y_U^\dagger ) + 2 (Y_U Y_U^\dagger) m_{H_u}^2   \nonumber \\
&  + (Y_DY_D^\dagger m_Q^2)+ ( m_Q^2Y_DY_D^\dagger) + 2 (Y_D m_D^2 Y_D^\dagger ) + 2 (Y_D Y_D^\dagger) m_{H_d}^2\nonumber \\
& + \frac{1}{3} g_Y^2  Tr(Y m^2) - \frac{g_Y^2}{9} |M_1|^2 - 6 g_2^2 |M_2|^2 - \frac{32}{3} |M_3|^2 + 2 A_U A_U^\dagger + 2 A_D A_D^\dagger \nonumber \\
16\pi^2\frac{d}{dt} m^2_{U} =&  2(Y_U^\dagger Y_U m_U^2) + 2(m_U^2 Y_U^\dagger Y_U ) + 4 (Y_U^\dagger m_Q^2 Y_U ) + 4 (Y_U^\dagger Y_U) m_{H_u}^2  \nonumber \\
& - \frac{4}{3} g_Y^2  Tr(Y m^2) - \frac{32 g_Y^2}{9} |M_1|^2  - \frac{32}{3} |M_3|^2  + 4 A_U^\dagger A_U \nonumber \\
16\pi^2\frac{d}{dt} m^2_{D} =& 2(Y_D^\dagger Y_D m_D^2) + 2(m_D^2 Y_D^\dagger Y_D ) + 4 (Y_D^\dagger m_Q^2 Y_D ) + 4 (Y_D^\dagger Y_D) m_{H_d}^2  \nonumber \\
& + \frac{2}{3} g_Y^2  Tr(Y m^2)  - \frac{8 g_Y^2}{9} |M_1|^2  - \frac{32}{3} |M_3|^2 + 4 A_D^\dagger A_D\nonumber \\
16\pi^2\frac{d}{dt} m^2_{L} =& (Y_E Y_E^\dagger m_L^2) + (m_L^2 Y_E Y_E^\dagger) + 2 (Y_E m_E^2 Y_E^\dagger ) + 2 (Y_E Y_E^\dagger) m_{H_d}^2  \nonumber \\
& - g_Y^2  Tr(Y m^2)  - 2 g_Y^2 |M_1|^2 - 6 g_2^2 |M_2|^2 + 2 A_E A_E^\dagger \nonumber \\
16\pi^2\frac{d}{dt} m^2_{E} =& 2(Y_E^\dagger Y_E m_E^2) + 2(m_E^2 Y_E^\dagger Y_E ) + 4 (Y_E^\dagger m_L^2 Y_E ) + 4 (Y_E^\dagger Y_E) m_{H_d}^2  \nonumber \\
&+ 2 g_Y^2  Tr(Y m^2)  - 8 g_Y^2 |M_1|^2 + 4 A_E^\dagger A_E
\end{align}

The MSSM $A$-terms run as
\begin{align}
16\pi^2 \frac{d}{dt} A_U =&( \lambda_S^2 + 3\lambda_T^2) A_U + (2A_S \lambda_S^2 + 6 A_T\lambda_T^2)Y_U \nonumber \\
&+ [4 A_U Y_U^\dagger + 2 A_D Y_D^\dagger+ 6\tr(A_U Y_U^\dagger)-\frac{26}{9} g_Y^2 M_1 - 6 g_2^2 M_2 - \frac{32}{3} g_3^2 M_3]Y_U  \nonumber \\
&+ [  Y_D Y_D^\dagger + 5 Y_U Y_U^\dagger+  3 \mathrm{tr}(Y_U^\dagger Y_U) - \frac{16}{3} g_3^2  - 3g_2^2 - \frac{13}{9} g_Y^2] A_U  \nonumber \\
16\pi^2 \frac{d}{dt} A_D =&( \lambda_S^2 + 3\lambda_T^2) A_D + (2A_S \lambda_S^2 + 6 A_T\lambda_T^2)Y_D \nonumber \\
&+ \bigg[ Y_U Y_U^\dagger + 5 Y_D Y_D^\dagger + 3\tr(Y_D Y_D^\dagger) + \tr(Y_E Y_E^\dagger) - \frac{7}{9} g_Y^2 - 3 g_2^2 - \frac{16}{3} g_3^2 \bigg] A_D \nonumber \\
&+ \bigg[ 2A_U Y_U^\dagger + 4 A_D Y_D^\dagger+ 6\tr(A_D Y_D^\dagger) + 2\tr(A_E Y_E^\dagger) \nonumber \\
&\qquad\qquad+ \frac{14}{9} g_Y^2 M_1 + 6 g_2^2 M_2 + \frac{32}{3} g_3^2 M_3 \bigg] Y_D \nonumber \\
16\pi^2 \frac{d}{dt} A_E=& (\lambda_S^2 + 3\lambda_T^2 ) A_E + (2A_S \lambda_S^2 + 6 A_T\lambda_T^2)Y_E \nonumber \\
&+ \bigg[ 5 Y_E Y_E^\dagger + 3\tr(Y_D Y_D^\dagger) + \tr(Y_E Y_E^\dagger) - 3 g_Y^2 - 3 g_2^2 \bigg] A_E \nonumber \\
&+ \bigg[ 4 A_E Y_E^\dagger +6\tr(A_D Y_D^\dagger) + 2\tr(A_E Y_E^\dagger) + 6 g_Y^2 M_1 + 6 g_2^2 M_2\bigg] Y_E 
\end{align}

The new $A$-terms run as
\begin{align}
16\pi^2\frac{d}{dt}A_S=& 6 \tr(A_U Y_U^\dagger)+6\tr( A_D Y_D^\dagger) + 2\tr(A_E Y_E^\dagger) + 8 A_S \lambda_S^2 + 12 A_T\lambda_T^2 \nonumber \\
&+ 2g_Y^2 M_1 + 6 g_2^2 M_2+ 4A_{\kappa_S} \kappa_S^2\nonumber \\
16\pi^2\frac{d}{dt}A_T=&6 \tr(A_U Y_U^\dagger)+6\tr( A_D Y_D^\dagger) + 2\tr(A_E Y_E^\dagger) + 4 A_S \lambda_S^2 + 16 A_T\lambda_T^2 \nonumber \\
&+ 2g_Y^2 M_1 + 14 g_2^2 M_2\nonumber \\
16\pi^2 \frac{d}{dt} A_{\kappa_S} =& 12 A_S \lambda_S^2  + 12 \kappa_S^2 A_{\kappa_S}.
\end{align}

\end{document}